\def\spose#1{\hbox to 0pt{#1\hss}}
\def\approxlt{\mathrel{\spose{\lower 3pt\hbox{$\sim$}}
        \raise 2.0pt\hbox{$$<$$}}}
\def\approxgt{\mathrel{\spose{\lower 3pt\hbox{$\sim$}}
        \raise 2.0pt\hbox{$>$}}}

\def\multleft#1{\hbox to size{\vbox {\halign {\lft{##}\cr #1}}\hfill}\par}
\def\multright#1{\hbox to size{\vbox {\halign {\rt{##}\cr #1}}\hfill}\par}

\def\today{\ifcase\month\or January\or February\or March\or April\or May\or
      June\or July\or August\or September\or October\or November\or December\fi
      \space\number\day, \number\year}
\def\$<${\thinspace}
\def\s{\hbox{\phantom{5}}}      

\def\boxit#1{\vbox{\hrule\hbox{\vrule\kern3pt\vbox{\kern3pt
          #1 \kern3pt}\kern3pt\vrule}\hrule}}

\def\cm{{\rm\thinspace cm}}

\def\erg{{\rm\thinspace erg}}

\def\keV{{\rm\thinspace keV}}

\def\ph{{\rm\thinspace ph}}
\def\s{{\rm\thinspace s}}

\def\ergpcmsqps{\hbox{$\erg\cm^{-2}\s^{-1}\,$}}

\def\phpcmsqps{\hbox{$\ph\cm^{-2}\s^{-1}\,$}}

\documentstyle[psfig,times]{mn}
\begin{document}
\hsize=6truein

\def\simless{\mathbin{\lower 3pt\hbox
   {$\rlap{\raise 5pt\hbox{$\char'074$}}\mathchar"7218$}}}   
\def\simgreat{\mathbin{\lower 3pt\hbox
   {$\rlap{\raise 5pt\hbox{$\char'076$}}\mathchar"7218$}}}   
\def\anisotropy{\frac{\Delta T}{T}}

\def\approxlt{\mathrel{\spose{\lower 3pt\hbox{$\sim$}}
        \raise 2.0pt\hbox{$<$}}}
\def\approxgt{\mathrel{\spose{\lower 3pt\hbox{$\sim$}}
        \raise 2.0pt\hbox{$>$}}}

\title{The X-ray variability of the Seyfert~1 galaxy MCG$-$6-30-15 from long {\it ASCA } and {\it RXTE } observations}

\author[J.~C.~Lee et al.]
{\parbox[]{6.in} {Julia~C.~Lee,$^1$\thanks {Current Address: MIT Center for Space Research; 77 Massachusetts Ave., NE80-6091; Cambridge MA 02139 USA} A.C.~Fabian,$^1$ Christopher~S.~Reynolds,$^2$ W.N.~Brandt,$^3$ and Kazushi Iwasawa$^1$ \\
\footnotesize \it $^1$ Institute of Astronomy; Madingley Road; Cambridge CB3 0HA \\
\it $^2$ Hubble Fellow : JILA; Campus Box 440; University of Colorado; Boulder, CO 80309-0440 USA \\ 
\it $^3$ Department of Astronomy and Astrophysics; The Pennsylvania State University; 525 Davey Lab; University Park, PA 16802 USA 
}}

\maketitle

\begin{abstract}
We present an analysis of the long Rossi X-ray Timing Explorer ({\it RXTE })
observation of the Seyfert~1 galaxy MCG$-$6-30-15, taken in July 1997.
We have earlier used the data to place constraints for the first time
on the iron abundance - reflection fraction relationship, and now
expand the analysis to investigate in detail the spectral X-ray
variability of the object. Our results show that the behaviour is
complicated. We find clear evidence from colour ratios and direct
spectral fitting that changes to the intrinsic photon index are taking
place. In general, spectral hardening is evident during periods of
diminished intensity, and in particular, a general trend for harder
spectra is seen in the period following the hardest {\it RXTE } flare.
Flux-correlated studies further show that the 3$-$10~keV photon index
$\Gamma_{3-10}$ steepens while that in the 10--20~keV band,
$\Gamma_{10-20},$ flattens with flux. The largest changes come from
the spectral index below 10~keV; however, changes in the intrinsic
power law slope (shown by changes in $\Gamma_{3-10}$), and reflection
(shown by changes in $\Gamma_{10-20}$) both contribute in varying
degrees to the overall spectral variability. We find that the iron
line flux $F_{\rm K\alpha}$ is consistent with being constant over
large time intervals on the order of days (although the {\it ASCA } and {\it RXTE }
spectra show that $F_{\rm K\alpha}$ changes on shorter time intervals
of order $\approxlt$ 10~ks), and equivalent width which anticorrelates
with the continuum flux, and reflection fraction. 
A possible interpretation for the iron line flux constancy
and the relative Compton reflection increase with flux from the 
flux-correlated data is an increasing ionization of the emitting disk surface,  
while spectral analysis of short time intervals surrounding flare events 
hint tentatively at observed spectral responses to the flare.
We present a simple model for partial ionization where the bulk of the
variability comes from within $6 r_g$.  Temporal analysis further
provides evidence for possible time ($\approxlt$ 1000s) and phase
($\phi \sim 0.6$ rad) lags. Finally, we report an apparent break in
the power density spectrum ($\sim 4-5
\times 10^{-6}$ Hz) and a possible 33 hour period.  Estimates for the
mass of the black hole in MCG$-$6-30-15 are discussed in the context of
spectral and temporal findings.

\end{abstract}


\section{INTRODUCTION}
Observational and theoretical progress over the past decade have
greatly improved our understanding of accreting black holes in Seyfert
galaxies. Time-averaged X-ray spectra reveal a hard power-law
continuum with a broad iron line and continuum reflection components
(Pounds et al 1990; Nandra \& Pounds 1994; Tanaka et al 1995; Nandra
et al 1997). The power-law emission is produced by thermal
Comptonization in $\sim 100$~keV hot gas (Zdziarski et al 1994) above
a thin accretion disc, which causes the reflection (Guilbert \& Rees
1988; Lightman \& White 1988; George \& Fabian 1991; Matt, Perola \&
Piro 1991). 

Spectral variability in some objects show that the emission is produced in flarelike
events which may move about over the inner accretion flow. The
information from such variability is complex and its interpretation is
not necessarily straightforward. Barr
\& Mushotzky (1986) and Wandel \& Mushotzky (1986), by invoking the
criterion for the fastest doubling time found in observations of AGN,
demonstrate that a correlation between the X-ray luminosity and
variability time scales exist and use this to place upper limits on
the black hole masses in their sample of AGN observed with a slew of
X-ray telescopes that include Ariel V, HEAO-1, and the Einstein
observatory.  Such a method however has been criticized for its
dependence on data quality and coverage (Lawrence et al. 1987; Mchardy
\& Czerny 1987). It does nevertheless provide a strong limit on the
size of the emitting region.  For example, Reynolds et al. (1995) and
Yaqoob et al. (1997) have reported a factor of 1.5 increase in flux in
as little as 100s in MCG$-$6-30-15.

Others have resorted to employing power density spectrum (PDS)
techniques (and variants thereof) as potential black hole mass
estimators (e.g.  Edelson \& Nandra 1999 for NGC3516; Nowak \& Chiang
1999 for MCG$-$6-30-15; Hayashida et al. 1998 introduces the {\it
normalized power spectrum density (NPSD)} for their sample of AGN
observed with {\it Ginga}).  However, even this is not without its
caveats. X-ray variability studies of AGNs thus far (with the
exception of IRAS18325-5926; Iwasawa et al. 1998) has shown that the
power-law spectrum is essentially chaotic with no characteristic
period.  The problem of unevenly sampled data streams, characteristic of X-ray
observations of AGN, is overcome by the solution of Lomb (1976) and Scargle
(1982; see Press et al 1992).

Others still have searched for time lags and leads using cross
correlation function techniques.  However, the problem associated with
unevenly sampled data is still a concern, and has been addressed by
e.g. Edelson \& Krolik (1988), Yaqoob et al (1997), Reynolds (1999),
and in this paper.

MCG$-$6-30-15 is a bright nearby ({\it z=0.0078}) Seyfert~1 galaxy that
has been extensively studied by every major X-ray observatory since
its identification.  A recent study that takes advantage of the high
energy and broad-band coverage of {\it RXTE } by Lee et al. (1998, 1999) with
respectively a 50~ks observation in 1996, and 400~ks 1997 observation
have confirmed the clear existence of a broad iron line and reflection
component in this object.  A previous study by Iwasawa et al. (1996) 
(hereafter I96) using a 4.5~day
observation with {\it ASCA } revealed the iron line profile to be variable,
which is confirmed in this long 1997 {\it ASCA } observation (Iwasawa et al. 1999,
hereafter I99) which was
observed simultaneously with {\it RXTE }.  Additionally, Lee et al. (1999)
have been able to constrain the iron abundance - reflection fraction
relation for the first time using the {\it RXTE } observation. Guainazzi et
al. (1999) give good bounds on the high energy cutoff of the continuum
from a {\it BeppoSAX } observation of MCG$-$6-30-15.

We investigate in this paper changes in the direct and reflected
components with our simultaneous {\it RXTE } and {\it ASCA } observations spanning a
good time interval of $\approx$ 400~ks, in order to speculate on possible
causes for variability. In Section~3, we investigate rapid variability
with colour ratios. This is followed in Section~4 by flux-correlated
studies, and a detailed analysis of the time intervals surrounding
the bright {\it ASCA } and {\it RXTE } flare shown in Fig.~\ref{fig1-ascaxteltc}.  In Section~5, we
shift gears to temporal studies and search (using cross-correlation
techniques) for the presence of time lags and leads, in order to
better understand the physical processes  that may give rise to each
other.  We discuss in Section~6 tentative evidence for the  presence
of a possible 33 hour period, and break in the power spectrum.
Finally, we discuss the spectral and temporal findings in Section~7.
In particular, we discuss the challenge that spectral findings may
pose  to current models for reflection by cold material,  and
implications from temporal studies for placing constraints on the size
of the emitting region and mass of the black hole in MCG$-$6-30-15. We
present also a simple model to explain some of our
enigmatic spectral findings.  We conclude in Section~8 with a summary
of the pertinent findings from this study.
 
\section{Observations}
MCG$-$6-30-15 was observed by {\it RXTE } over the period from 1997 Aug 4
to 1997 Aug 12 by both the Proportional Counter Array (PCA) and
High-Energy X-ray Timing Experiment (HEXTE) instruments.  
We note that good data covered $\sim$400~ks, even though the {\it RXTE } observation 
spanned $\sim$700~ks. It was 
simultaneously observed for $\rm \sim 200~ks$ by the {\it ASCA } Solid-state
Imaging  Spectrometers (SIS) over the period 1997 August 3 to 1997
August 10 with a half-day gap  part way through the observation.  We
concentrate only on the {\it RXTE } PCA observations in this paper.

\subsection{Data Analysis}
We extract PCA light curves and spectra from only the top Xenon layer
using the {\sc ftools 4.2} software. The light curves
were extracted with the default 16s time bins for Standard 2
PCA data time resolution.   Data from PCUs 0, 1,
and 2 are combined to improve signal-to-noise at the expense of
slightly blurring the spectral  resolution.  Data from the remaining
PCUs (PCU 3 and 4) are excluded because these instruments are known to
periodically suffer discharge and are hence  sometimes turned off.

Good time intervals were selected to exclude any earth or South
Atlantic Anomaly (SAA) passage occulations, and to ensure stable
pointing.  We also filter out electron contamination events.

We generate background data using {\sc pcabackest v2.0c} in order  to
estimate the internal background caused by interactions between the
radiation/particles and the detector/spacecraft at the time of
observation.  This is done by matching the conditions of observations
with those in various model files.  The model files used are the
L7-240 background models which are intended to be specialized for
application to faint sources with count rate less than 40~$\rm cts\,s^{-1}\,PCU^{-1}$.

The PCA response matrix for the {\it RXTE } data set was created using {\sc
pcarsp  v2.36}.  Background models and response matrices are
representative of the most up-to-date PCA calibrations.

\begin{figure*}
\psfig{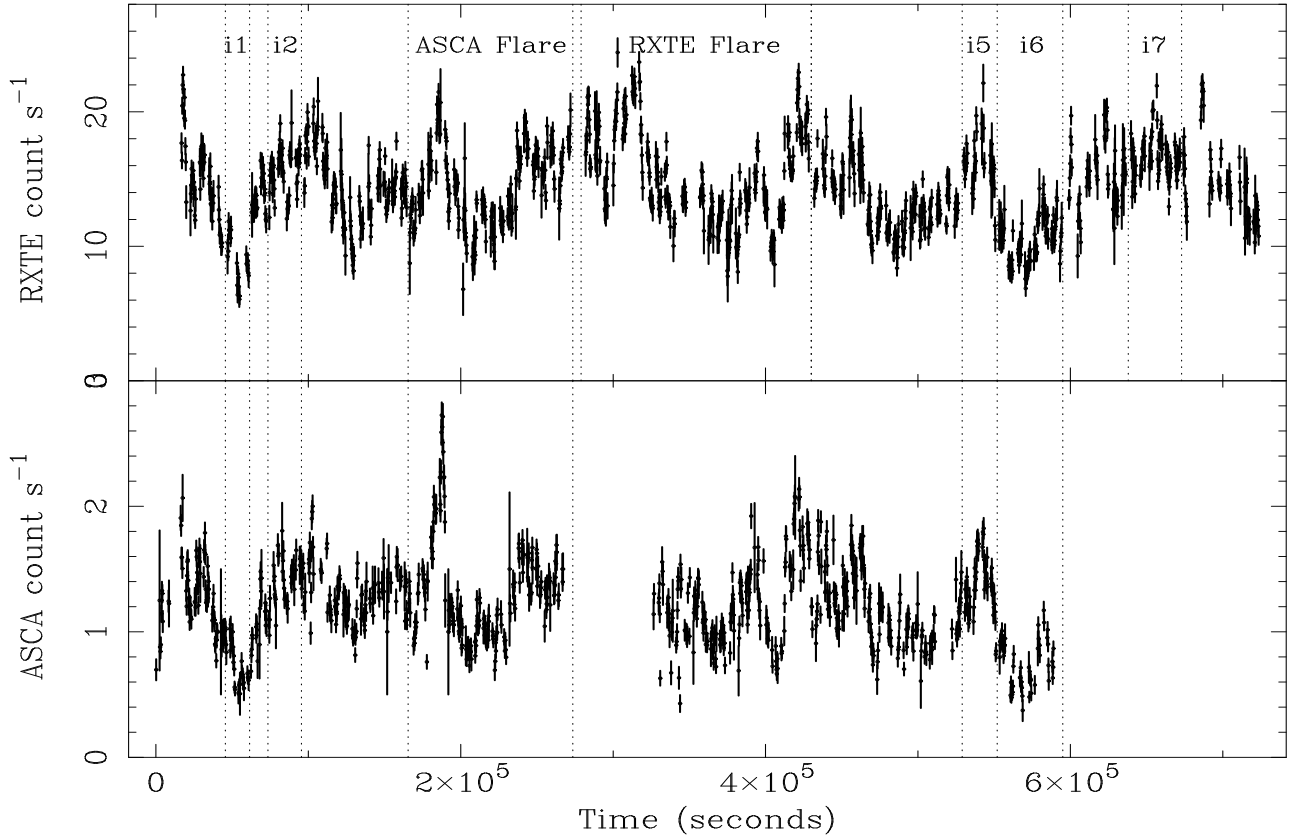}
\caption[h]{Background-subtracted light curve of MCG$-$6-30-15 for observations with 3 PCUs in the 2-60~keV band for the {\it RXTE } PCA, and 0.6$-$10~keV {\it ASCA } SIS. The epoch of the start and stop times for {\it RXTE} and {\it ASCA } respectively is 1997 August 4, 03:27:06 (UT) and 1997 August 12, 12:34:14 (UT) ({\it RXTE }), and 1997 August 3 to 1997 August 10 ({\it ASCA }). Light curves are binned using 256s time intervals. The intervals corresponding to the {\it ASCA }  and {\it RXTE } flares are further subdivided and discussed in depth in Section~4.3. }
\label{fig1-ascaxteltc}
\end{figure*}

Figure~\ref{fig1-ascaxteltc} shows the background subtracted light curve for {\it ASCA } and
{\it RXTE }, with 256s binning.  The time intervals corresponding to the 
{\it ASCA } and {\it RXTE } flares are decomposed further and analyzed in depth in
Section~4.3.  Significant variability can be seen in both light curves on
short and long timescales. Flare and minima events are seen to
correlate temporally in both light curves.  

\section{Rapid Variability}
Intraday variability has been seen in most quasars and AGNs; in
MCG$-$6-30-15, rapid X-ray variability  on the order of 100~s has been
reported by a number of workers (e.g. Reynolds et al. 1995; Yaqoob et
al. 1997).  These significant luminosity changes on time scales as
short as minutes can have strong implications for restricting the size
of the emitting region, and efficiency of the central engine.   Colour
ratios are good tools for discerning the processes that may give rise
to these rapid variability effects.  For the purposes of our study, we
define the following count rate hardness ratios in order to assess the
processes that may give rise to variability :

\begin{equation}
R_1 = \frac{5-7~keV}{3-4.5~keV}
\end{equation}

\begin{equation}
R_2 = \frac{7.5-10~keV}{3-4.5~keV}
\qquad\mbox{, }\qquad R_3 = \frac{7.5-10~keV}{5-7~keV}
\end{equation}

\begin{equation}
R_4 = \frac{10-20~keV}{3-4.5~keV}
\qquad\mbox{, }\qquad R_5 = \frac{10-20~keV}{5-7~keV}
\end{equation}
\[
\qquad\mbox{}\qquad  R_6 = \frac{10-20~keV}{7.5-10~keV}
\]
We expect the reflection component to dominate above 10~keV.  At the
lower energies, the dominant features include the iron line between
5.5-6.5~keV and the warm absorber below 2~keV.   We note also that the
iron line contribution to the overall flux in the 5-7~keV band is only
$\sim$ 15\%, and hence be hereafter referred to as the iron line
band. 

\begin{table}
\begin{center}
\begin{tabular}{|c|c|}
\multicolumn{2}{c} {}\\
{\em \rm Hardness ratio} & {\em $\chi_\nu^2$ } \\
\hline
\hline
R1 & 3.7\\
R2 & 3.25\\
R3 & 0.95\\
R4 & 4.42\\
R5 & 1.95\\
R6 & 1.24\\
\hline
\hline

\end{tabular}
\end{center}
\caption{Assessment for amount/absence of variability for the hardness ratios shown
on the left panels of Figs.~\ref{fig2-hratio}$-$\ref{fig4-hratio}.  $\chi_\nu^2$ is 
evaluated for 138 degrees of freedom. }
\label{tab-hardnessvar}
\end{table}

The $\rm R_1$ assessment of the (5-7~keV) iron line band versus (3-4.5~keV) 
lower continuum (Fig.~\ref{fig2-hratio}a) shows that the source is intrinsically
harder during the minima.  Similarly, $\rm R_2$ hardness ratio of the
(7.5-10~keV) upper continuum to the (3-4.5~keV) lower continuum
reveals a similar trend (Fig.~\ref{fig3-hratio}a); likewise, $\rm R_4$ assessment of
the (10-20~keV)  reflection hump and (3-4.5~keV) lower continuum
(Fig.~\ref{fig4-hratio}a).  $\rm R_3$, $\rm R_5$, $\rm R_6$ show no obvious trends
(respectively Figs.~\ref{fig3-hratio}b,\ref{fig4-hratio}b,\ref{fig4-hratio}c) from 
count-rate versus hardness ratio plots, although subtle correlations can be seen in the time versus
hardness ratio plots.  In particular Figs.~\ref{fig4-hratio}b and \ref{fig4-hratio}c show dramatic
hardening of the spectrum during the time periods following the 
{\it RXTE } flare.  (We mark the beginning of the {\it ASCA } and {\it RXTE } flares
respectively with `A' and `X' in the light curves shown in Figs.~\ref{fig2-hratio}$-$\ref{fig4-hratio}.)
In order to further quantify the amount / absence of variablity, we apply a $\chi^2$ test
to the hardness ratio trends as shown in the left panels of Figs.~\ref{fig2-hratio}$-$\ref{fig4-hratio}.
Table~\ref{tab-hardnessvar} details these results for 138 degrees of freedom. 

\subsection{The unusual properties of the deep minimum period following the hard {\it RXTE } flare}
\begin{figure*}
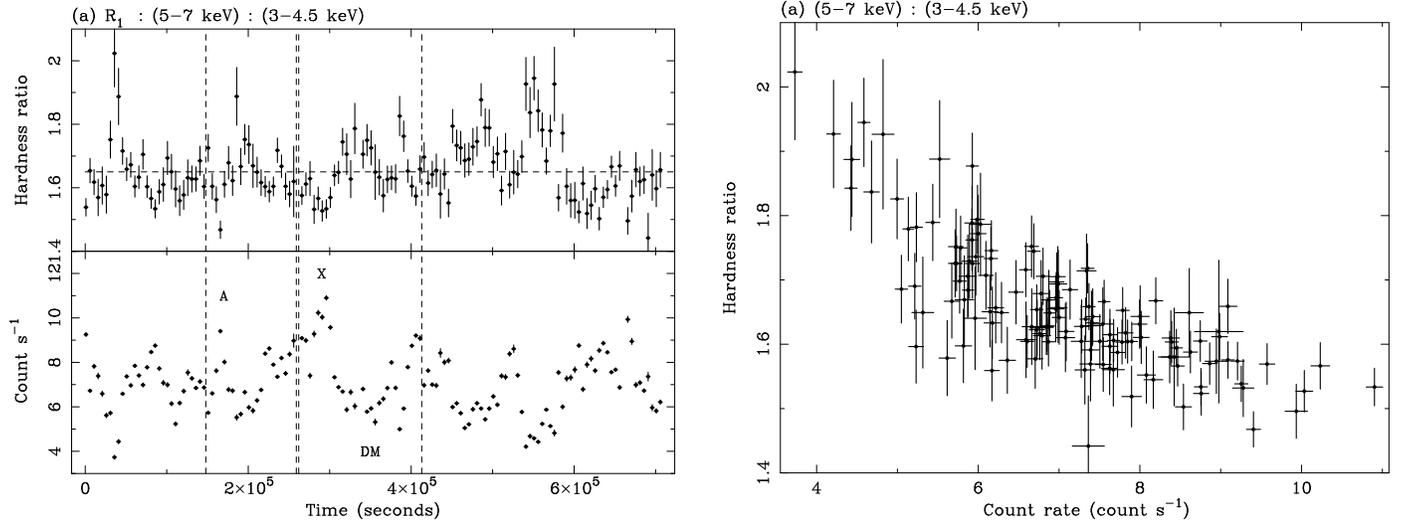

  \hbox{
    \psfig{figure=m762f2a1.ps,width=0.5\textwidth,angle=270}
    \hspace{0.5cm}
    \psfig{figure=m762f2a2.ps,width=0.5\textwidth,angle=270} }
\caption[h]{Hardness ratio plotted against time and countrate for
the iron line band (5-7~keV) versus lower continuum (3-4.5~keV). Light curves are 
binned into 5000s intervals, and represent the sum of the light curve of the two energy bands
being compared.}
\label{fig2-hratio}
\end{figure*}

\begin{figure*}
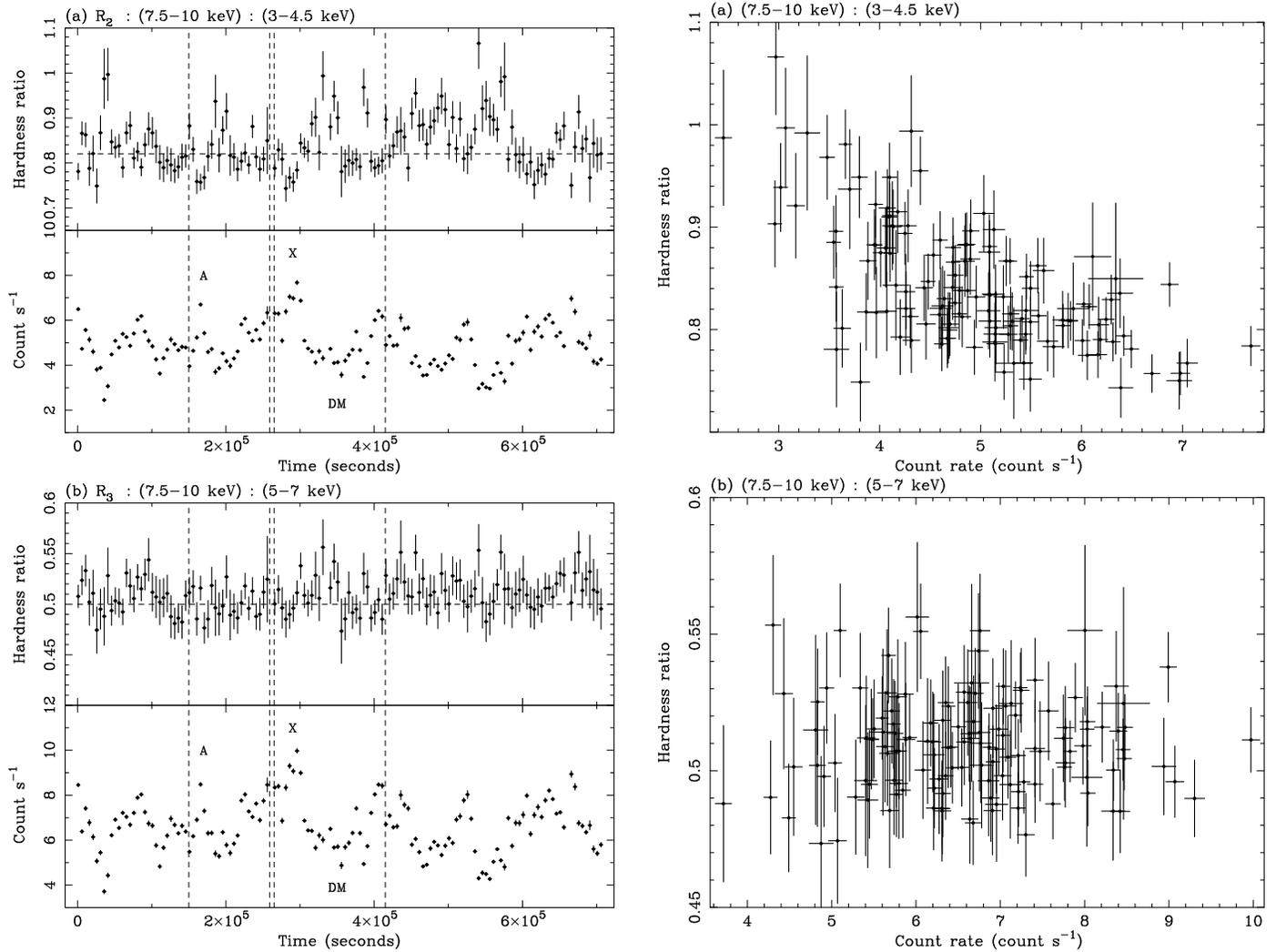

  \hbox{ \psfig{figure=m762f3a1.ps,width=0.5\textwidth,angle=270}
    \hspace{0.5cm}
    \psfig{figure=m762f3a2.ps,width=0.5\textwidth,angle=270} }
  \hbox{ \psfig{figure=m762f3b1.ps,width=0.5\textwidth,angle=270}
    \hspace{0.5cm}
    \psfig{figure=m762f3b2.ps,width=0.5\textwidth,angle=270} }
\caption[h]{Hardness ratio plots against time and count rate of the upper continuum (7.5-10~keV) versus (a) lower continuum (3-4.5~keV) and (b) iron line band (5-7~keV). Light curves are binned into 5000s intervals, and represent the sum of the light curve of the two energy bands
being compared..}
\label{fig3-hratio}
\end{figure*}

\begin{figure*}
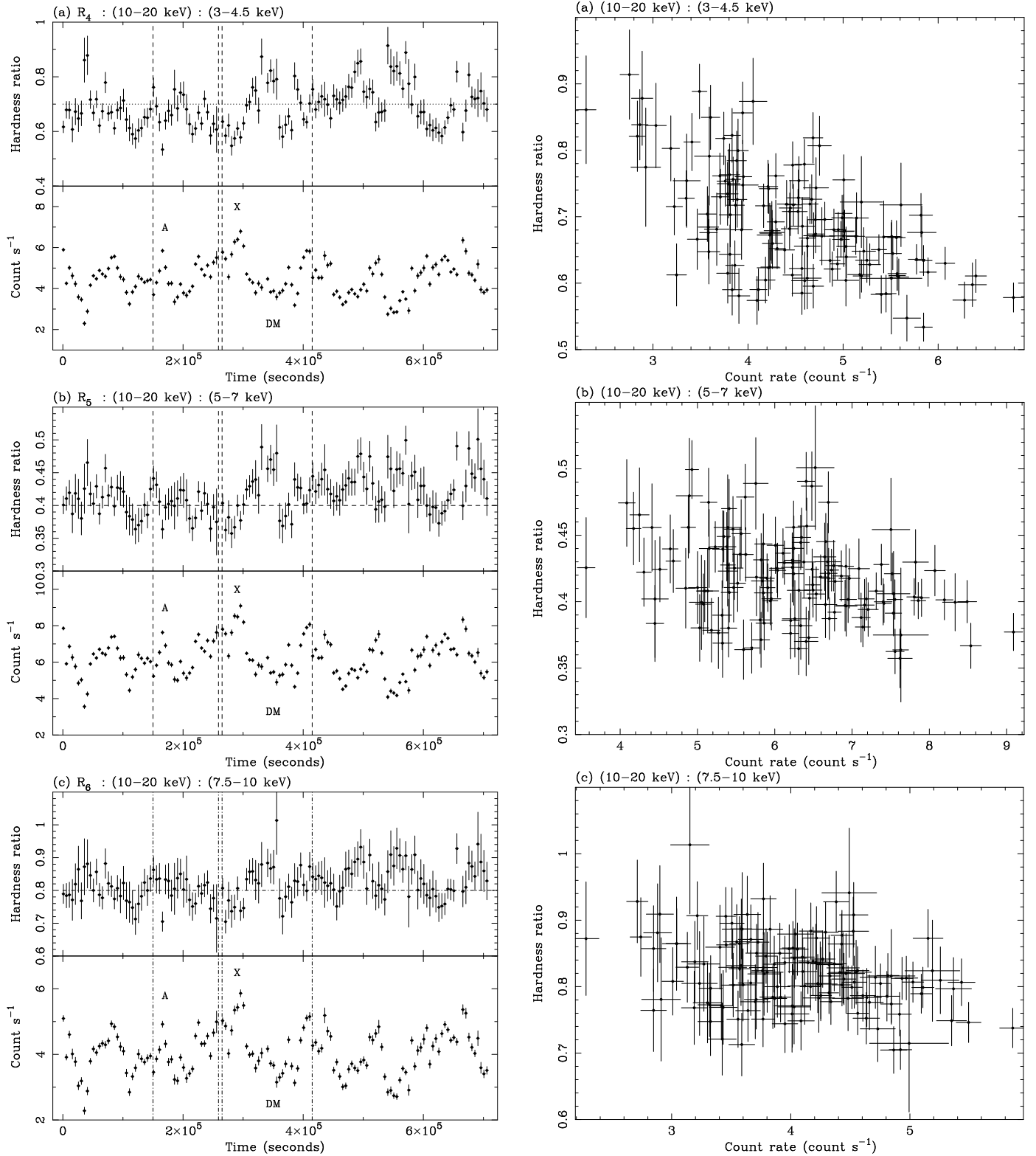

  \hbox{ \psfig{figure=m762f4a1.ps,width=0.50\textwidth,angle=270}
    \hspace{0.5cm}
    \psfig{figure=m762f4a2.ps,width=0.50\textwidth,angle=270} }
  \hbox{ \psfig{figure=m762f4b1.ps,width=0.5\textwidth,angle=270}
    \hspace{0.5cm}
    \psfig{figure=m762f4b2.ps,width=0.5\textwidth,angle=270} }
  \hbox{ \psfig{figure=m762f4c1.ps,width=0.5\textwidth,angle=270}
    \hspace{0.5cm}
    \psfig{figure=m762f4c2.ps,width=0.5\textwidth,angle=270} }
\caption[h]{Hardness ratio plotted against time and count rate for the reflection hump (10-20~keV) versus (a) the lower continuum (3-4.5~keV), (b) iron line band (5-7~keV), and (c) upper continuum (7.5-10~keV). Light curves are binned into 5000s intervals, and represent the sum of the light curve of the two energy bands being compared..}

\label{fig4-hratio}
\end{figure*}

We mark the regions corresponding to the {\it ASCA } and {\it RXTE } flare events in
Figs.~\ref{fig2-hratio}$-$\ref{fig4-hratio}. The time span corresponding to the {\it ASCA } and {\it RXTE } flares were chosen
in order that the times surrounding the flares and minimum (following said flare)
can be compared. The spectral behaviour during these periods is
thoroughly investigated in Section~4.3.  For present purposes,  we
draw the reader's attention to the deep minimum immediately following
the bright {\it RXTE } flare. (The general location of this minimum in the light
curve is labeled with `DM' in Figs.~\ref{fig2-hratio}$-$\ref{fig4-hratio}.) Despite the hardness ratio results discussed
previously (e.g. no obvious trends seen in $\rm R_3$, $\rm R_5$, $\rm R_6$),
 the spectrum during this time period always hardens in all
colours, whereas this is not necessarily the case with the other
minima events (e.g. the  deep minimum following the soft {\it ASCA }
flare). In addition, it is not clear whether the spectrum reaches its lowest
minimum slightly before or after an observed hardening.

We also bring attention (by means of a horizontal line through the 
hardness ratio versus time plots in Figs.~\ref{fig2-hratio}$-$\ref{fig4-hratio}) to the general
trend for harder spectra in the $\sim$ 260~ks time period 
that begins with the {\it RXTE } flare.

\subsection{Possible causes of Hardness Ratio Variability}
Many complicated processes contribute to the overall temporal and
spectral behaviour of an AGN.  It is necessary to disentangle these
components in order to assess the physical processes that are
responsible for the observed variability phenomenon.  Effects
due to reprocessing are significant only at the higher energies, the
(10-20~keV) band being the most sensitive probe of any subtle changes
in the reflection component.  Martocchia \& Matt (1996) suggested
that gravitational bending/focusing as the X-ray source gets closer to
the black hole will enhance the amount of reflection.  Therefore, it
is possible that flares closer to the hole during the minima can
enhance the amount of reflection, through increased beaming of the 
emission towards the disk.  We conclude however from the combined
hardness ratio findings described thus far that the spectral hardening during
periods of diminished intensity point largely to changes in the
intrinsic photon index $\Gamma$ being the main culprit for the
observed spectral variability. We are led to this conclusion because
spectral changes are seen in all bands, which is most likely to be due to
changes in the spectral index.  In other words, we may be seeing the
effects of changes in the  coronal temperature affecting the
intrinsic power law slope, or that  changes in $\Gamma$ are due to
coverage of flares such that we are observing the effects of flares
occurring at different heights (e.g. Poutanen \& Fabian 1999).  
However, as we shall now show,
direct spectral analysis reveals a more complex situation.

\section{Spectral Fitting}
Having established that variability effects are many, 
we next investigate spectral changes in time sequence in
order to obtain a better understanding of variability phenomena
between the different flux states with particular emphasis on the
flares and deep minima.  Data analysis is restricted to the 3
to 20~keV PCA energy band.  The lower energy restriction at 3~keV 
is selected in order that the necessity for modelling
photoelectric absorption due to Galactic ISM material, or the warm
absorber that are known to be present in this object is removed. (Lee
et al. 1999 have shown that effects due to the warm absorber 
at 3~keV are negligible for this data set.) 

\subsection{Spectra selected in time sequence and by flux}
We choose the time intervals (Fig.~\ref{fig1-ascaxteltc}) in order to contrast and study the
different variability states. They are defined such that intervals
{\bf i1} and {\bf i6} correspond to the two {\it RXTE } deep minima, and
intervals {\bf i2} and {\bf i7} for the relatively calm periods
following these minima, and {\bf i5} a flare event for comparison.
We investigate in detail in Section~4.3 the times surrounding the {\it RXTE }
counterpart of the soft {\it ASCA } flare, and the
hard flare observed by {\it RXTE } (unfortunately missed by {\it ASCA }), as depicted in Fig.~\ref{fig1-ascaxteltc}
and subsequently in Figs~\ref{fig10-ascaflare}, and \ref{fig12-xteflare}.  Both the  
{\it ASCA } and {\it RXTE } light curves for the full observation are shown in
Fig.~\ref{fig1-ascaxteltc}.

We also separate the data according to flux in order to assess whether
a clear picture can be developed of the dominant processes that may be
at work for a given flux state.  To do so, we separate the 400~ks
observation by flux, with {\bf f2} and {\bf f3} being the intermediate
states between the lowest {\bf f1} ($\rm 2.84 \times 10^{-11}$
\ergpcmsqps), and highest {\bf f4} ($\rm 4.79 \times 10^{-11}$
\ergpcmsqps) fluxes.  Properties of these flux levels (in the 2-60~keV, 
and 3-20~keV energy bands), and  {\bf i1}$-${\bf i7}
intervals are detailed respectively in Table~\ref{tab1-pl}, and Table~\ref{tab5-intproperties}.

\subsection{Temporal Variability and Spectral Features }
A nominal fit to the entire data set (ie. {\it ASCA }, {\it RXTE } PCA and HEXTE)
demonstrated the clear existence of a redshifted broad iron K$\alpha$
line at $\sim$ 6.0~keV and reflection hump between 20 and 30~keV as
shown in Lee et al. (1998, 1999).  In general, fits to the solar
abundance models of George \& Fabian 1991 (hereafter GF91) are better
than simple power law fits and further reinforce the preference for a
reflection component.  Since the reflected continuum does not
contribute significant flux to the observed spectrum below 10~keV, a
simple power law and iron line fit to the data below 10~keV will reveal
changes in the  intrinsic power-law of the X-ray source. If indeed
spectral variability is due to changes in the intrinsic power law
slope which would implicate changes in the conditions of the X-ray
emitting corona, we should see noticeable changes in the values of
$\Gamma_{(3-10)}$ between the different temporal states. However,
because reflection only significantly affects energies above 10~keV,
changes seen in $\Gamma_{(10-20)}$ would suggest that variability may
be due to the amount of reprocessing.  (We have tested this premise by 
comparing simulated spectra in {\sc xspec} using the {\sc pexrav} model and find that the overall flux contribution 
from reflection alone is $>$~60 per cent above 10~keV.) This would have strong
implications for geometry (ie. where the direct X-ray flares are
partially obscured), motion of the source (eg. Reynolds \& Fabian
1997; Beloborodov 1998), or gravity (light-bending effects that will
beam/focus more of the emission down towards the disk; Martocchia \&
Matt 1996).

\subsubsection{The different flux states}
\begin{figure}
\centerline{\psfig{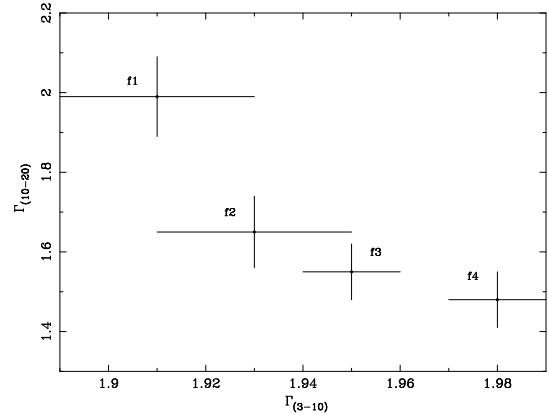}}
\caption[h]{Intrinsic photon index $\Gamma_{3-10}$ vs $\Gamma_{10-20}$ of 
reflection component shows that changes in both are evident.}
\label{fig5-gammagamma}
\end{figure}

We first investigate whether a correlation exists between flux and the
various fit parameters, and in particular whether changes in
reflection are dramatic.  We do so by first fitting a simple model
that consists of a power law and redshifted  Gaussian to the 3-10~keV
band, and comparing that with a power law fit to the 10-20~keV band.
In doing so, we find that while the intrinsic 3-10~keV power law slope
increases, $\Gamma_{10-20}$ appears to flatten with increasing flux (Fig.~\ref{fig5-gammagamma}).
Additionally, the iron $\rm K\alpha$ flux $F_{\rm K\alpha}$ does not
change with any statistical significance while the flux between the
lowest {\bf f1} and highest {\bf f4} flux nearly doubles (Table~\ref{tab1-pl}); similar
results for the constancy of the iron line in MCG$-$6-30-15 was noted
by McHardy et al. (1998), and by Chiang et al. (1999)
for Seyfert~1 galaxy NGC5548.  The ratio
plot of data-to-model using a simple power law fit to the  3-20~keV
data clearly illustrates the difference in the line and reflection
component between the lowest and highest flux states in Fig.~\ref{fig7-ratioplts}a.

\begin{table*}
\begin{center}
\begin{tabular}{|c|c|c|c|c|c|c|c|c|c|}
\multicolumn{10}{c}{\sc Change in $\Gamma$ for different states of MCG$-$6-30-15} \\
\hline
{\em \rm Flux} & {\em $\rm \Gamma_{3-10}^a$} & {\em $F_{\rm K\alpha}^b$} & {\em $\rm \Gamma_{10-20}^c$} & {\em $W_{\rm K\alpha}^d$ } & {\em $\rm Flux ^e$ } & {\em \rm Exposure} & {\em \rm 2-60~keV}  & {\em \rm Flux cuts} & {\em \rm Time range}  \\
{\em \rm Levels}  && {\em \phpcmsqps}  & & {\em $\rm eV$} & {\em \rm (3$-$10~keV) } & {\em \rm $\rm 10^4$ s} & {\em $\rm ct$ $\rm s^{-1}$} & {\em $\rm ct$ $\rm s^{-1}$} & {\em $\rm 10^4 s$ } \\
\hline
\hline
f1 & 1.91 $\pm$ 0.02 & $1.81^{+0.16}_{-0.14}$ & 1.99 $\pm$ 0.10& $466^{+41}_{-36}$ & 2.84 & 7.8  & 9.89 $\pm$ 0.34 & f $<$ 12 & full \\
f2 & 1.93 $\pm$ 0.02 & $1.89^{+0.19}_{-0.17}$ & 1.65 $\pm$ 0.09& $399^{+39}_{-37}$ & 3.44 & 5.7  & 13.03 $\pm$ 0.41  & $\rm 12 \le f < 14$ & full\\
f3 & 1.95 $\pm$ 0.01 & $1.72^{+0.14}_{-0.14}$ & 1.55 $\pm$ 0.07& $313^{+25}_{-25}$ & 3.94 & 7.7  & 15.43 $\pm$ 0.36 & $\rm 14 \le f < 17$ & full\\
f4 & 1.98 $\pm$ 0.01 & $1.85^{+0.19}_{-0.17}$ & 1.48 $\pm$ 0.07& $272^{+28}_{-28}$ & 4.79 & 6.2 & 19.33 $\pm$ 0.40 & $\rm f \ge 17$ & full \\
\hline
\hline

\end{tabular}
\caption{Results are quoted from simple power law fits. 
$^a$ Power-law photon index for 3~keV~$<$~E~$<$~10~keV.  
$^b$ Flux of iron emission line in units of $10^{-4}$ \phpcmsqps .
$^c$ Power-law photon index for 10~keV~$<$~E~$<$~20~keV.  $^d$ Equivalent
width of the iron emission line.  $^e$ 3-10~keV flux in units of
$10^{-11}$ \ergpcmsqps
Intervals f1-f4 correspond to the full time-averaged
spectrum separated according to flux.  Count rates are
given for the 2-60~keV energy band of the {\it RXTE } PCA.  }

\label{tab1-pl}
\end{center}
\end{table*}

\begin{table*}
\begin{center}
\begin{tabular}{|c|c|c|c|c|c|c|c|c|c|}
\multicolumn{10}{c}{\sc Spectral Fits using power-law with reflection model for different flux states of MCG$-$6-30-15}\\
\hline
{\em \rm Data} & {\em $\rm ^a \Gamma_{4-20}$} & {\em $\rm ^b A$ } & {\em
$\rm ^c refl$ } & {\em $\rm ^d LineE$ } & {\em $\rm ^e \sigma$} & {\em $^f F_{\rm K\alpha}$} &
{\em $\rm ^g W$ (eV)} & {\em $\rm ^h Flux$ } & {\em $\rm ^i \chi^2$}  \\
\hline
\hline
f1 & $1.88^{+0.03}_{-0.03}$ & $1.16^{+0.03}_{-0.04}$ & $0.35^{+0.15}_{-0.16}$ & $6.04^{+0.05}_{-0.05}$ & $0.54^{+0.07}_{-0.06}$ & $1.50^{+0.11}_{-0.12}$ & $375^{+28}_{-30}$ & 4.61 & 76 \\
f2 & $1.97^{+0.03}_{-0.03}$ & $1.57^{+0.05}_{-0.05}$ & $0.91^{+0.22}_{-0.20}$ & $6.03^{+0.07}_{-0.07}$ & $0.54^{+0.09}_{-0.10}$ & $1.31^{+0.16}_{-0.16}$ & $261^{+31}_{-31}$ & 5.73 & 44 \\
f3 & $2.01^{+0.03}_{-0.02}$ & $1.91^{+0.05}_{-0.05}$ & $1.10^{+0.19}_{-0.16}$ & $6.05^{+0.06}_{-0.06}$ & $0.39^{+0.09}_{-0.10}$ & $0.96^{+0.12}_{-0.11}$ & $167^{+21}_{-19}$ & 6.58 & 34 \\
f4 & $2.07^{+0.03}_{-0.02}$ & $2.51^{+0.09}_{-0.05}$ & $1.37^{+0.23}_{-0.14}$ & $5.98^{+0.09}_{-0.10}$ & $0.36^{+0.14}_{-0.21}$ & $0.72^{+0.14}_{-0.19}$ & $98^{+19}_{-26}$ & 8.03 & 35 \\
\hline

\end{tabular}
\caption{$^a$ Power-law photon index.
$^b$ Power-law flux at 1~keV, in units of $10^{-3}$ $\ph\cm^{-2}\s^{-1}\keV^{-1},$
$^c$ Reflective fraction = $\Omega / 2\pi$.
$^d$ Energy of the iron $K\alpha$ emission line.
$^e$ Line width in units of eV.
$^f$ Flux of iron emission line in units of $10^{-4}$ \phpcmsqps .
$^g$ Equivalent width of the emission line.
$^h$ 3-20~keV flux in units of $10^{-11}$ \ergpcmsqps.
$^i$ $\chi^2$ for 39 degrees of freedom.}

\label{tab2-pexrav}
\end{center}
\end{table*}

\begin{table*}
\begin{center}
\begin{tabular}{|c|c|c|c|c|c|c|c|c|}
\multicolumn{9}{c}{\sc Spectral Fits using power-law with edge model for different flux states of MCG$-$6-30-15}\\
\hline
{\em \rm Data} & {\em $\rm ^a \Gamma_{3-10}$} & {\em $\rm ^b E_{\rm K\alpha}$ } & {\em $\rm ^c \sigma$} & {\em $^d F_{\rm K\alpha}$} & {\em $\rm ^e W_{\rm K\alpha}$ (eV)} & {\em $^f \rm E_{edge} (keV)$ } & {\em $ ^g \tau$} & {\em $\rm ^h \chi^2$}  \\
\hline
\hline
f1 & $1.83^{+0.03}_{-0.04}$ & $6.12^{+0.08}_{-0.07}$ & $0.65^{+0.11}_{-0.09}$ & $1.59^{+0.17}_{-0.20}$ & $406^{+43}_{-51}$ & $7.0 + 0.20$ & $0.13^{+0.06}_{-0.07}$ & 15  \\
f2 & $1.84^{+0.03}_{-0.04}$ & $6.16^{+0.11}_{-0.04}$ & $0.65^{+0.16}_{-0.11}$ & $1.56^{+0.43}_{-0.22}$ & $325^{+90}_{-46}$ & $7.0 + 0.15$ & $0.15^{+0.07}_{-0.05}$ & 23   \\
f3 & $1.89^{+0.03}_{-0.02}$ & $6.13^{+0.06}_{-0.06}$ & $0.53^{+0.09}_{-0.10}$ & $1.43^{+0.18}_{-0.20}$ & $259^{+33}_{-36}$ & $7.0 + 0.23$ & $0.10^{+0.40}_{-0.40}$ & 13  \\
f4 & $1.92^{+0.02}_{-0.02}$ & $6.10^{+0.08}_{-0.08}$ & $0.60^{+0.12}_{-0.13}$ & $1.46^{+0.43}_{-0.23}$ & $214^{+63}_{-34}$ & $7.0 + 0.16$ & $0.11^{+0.04}_{-0.03}$ & 13  \\
\hline

\end{tabular}
\caption{$^a$ Power-law photon index.
$^b$ Energy of the iron K$\alpha$ emission line.
$^c$ Line width in units of keV. 
$^d$ Flux of iron emission line in units of $10^{-4}$ \phpcmsqps .
$^e$ Equivalent width of the emission line (eV).
$^f$ Energy of absorption edge (redshifted to z=0.0078 appropriate for MCG$-$6-30-15) in units of keV.
$^g$ Maximum value for the optical depth at threhold energy.
$^h$ $\chi^2$ for 12 degrees of freedom.}

\label{tab-powithedge}
\end{center}
\end{table*}

Having demonstrated the existence of a strong reflection component in
Fig.~\ref{fig7-ratioplts}a, and assessed its nature with more complicated fits in Lee et
al. (1999) for this data set, we next investigate the features of
reflection in detail for the different fluxes by fitting the  data
with a multicomponent model that includes the reflected spectrum.  The
underlying continuum is fit with the model {\sc pexrav} which is a
power law with an  exponential cut off at high energies reflected by
an optically thick slab of  neutral material (Magdziarz \& Zdziarski
1995). We fix the inclination angle of the reflector at $30^\circ$ so
as to agree with the disk inclination one obtains when fitting
accretion disk models to the iron line profile as seen by {\it ASCA }
(Tanaka et al. 1995). Due to  the strong coupling between the fit
parameters of $\Gamma$, abundances, and reflection, we fix the  low-Z
and iron abundance respectively at 0.5 and 2 solar abundances as
determined by Lee et al. (1999) for the fits presented in Table~\ref{tab2-pexrav}; the
high energy cutoff is  fixed at 100~keV appropriate for this object
(Guainazzi et al. 1999; Lee et al. 1999).  An additional Gaussian
component is added to model the iron line.

\begin{figure*}
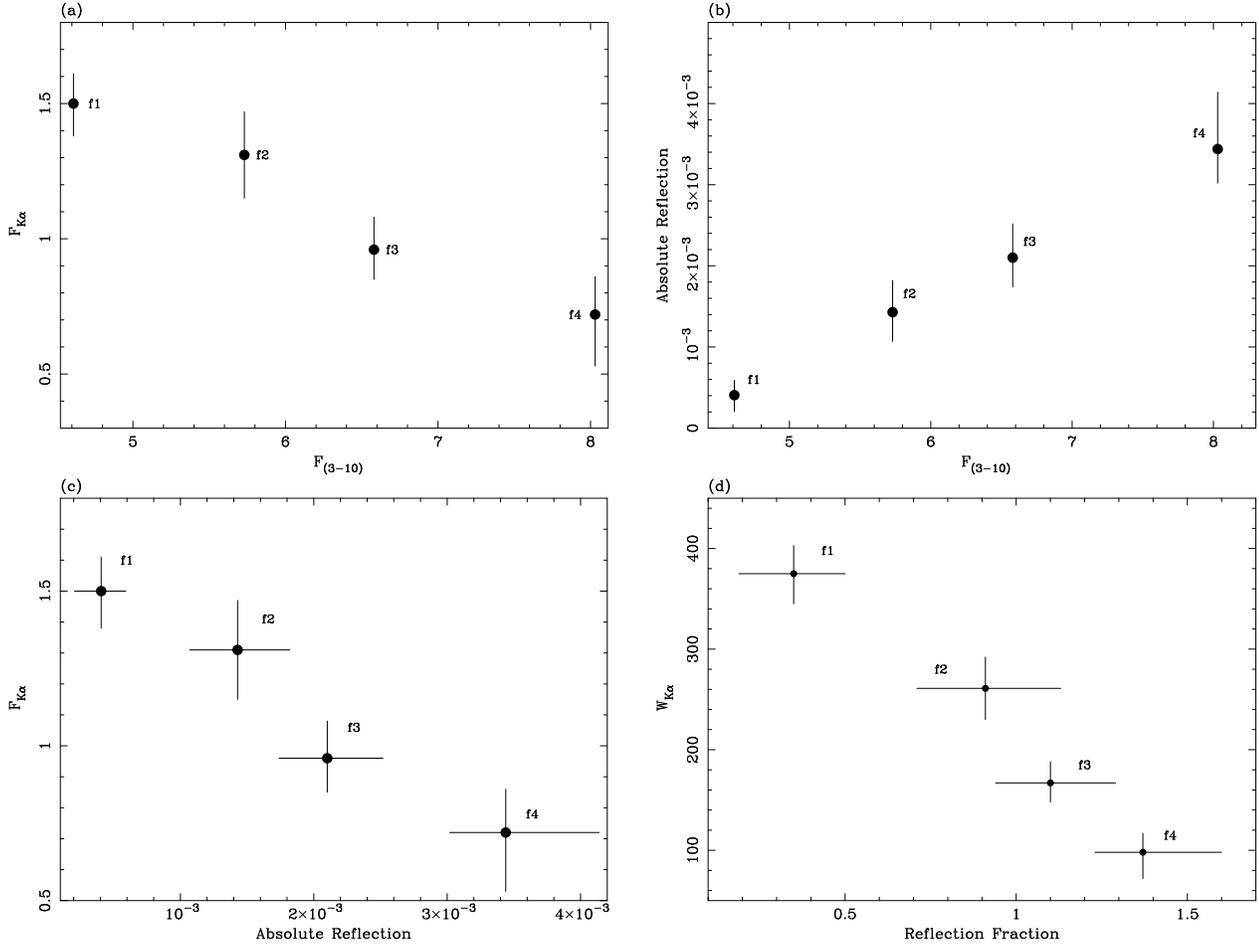

  \hbox{
    \psfig{figure=m762f6a.ps,width=0.45\textwidth,angle=270}
    \hspace{0.5cm}
    \psfig{figure=m762f6b.ps,width=0.45\textwidth,angle=270}} 
  \hbox{
    \psfig{figure=m762f6c.ps,width=0.45\textwidth,angle=270}
    \hspace{0.5cm}
    \psfig{figure=m762f6d.ps,width=0.45\textwidth,angle=270} }
\caption[h]{Plots representing results from complex fits using the model : {\sc pexrav} + {\sc gaussian} discussed in Section~4.2.1, and detailed in Table~\ref{tab2-pexrav}. 'Absolute reflection' refers to the absolute normalization of the reflection component (in units of $10^{-2}$ $\ph\cm^{-2}\s^{-1}\keV^{-1}$).}
\label{fig6-refewfe}
\end{figure*}

Using this complex model, we find that the 4-20~keV power law slope and reflection fraction
$R$ increases with flux (Table~\ref{tab2-pexrav}) while the strength of the iron line, 
$F_{\rm K\alpha}$ decreases (Fig.~\ref{fig6-refewfe}a), the latter in contrast to the findings for a
constant $F_{\rm K\alpha}$ discussed in the context of simpler
fits. ($F_{\rm K\alpha}$ is defined as the  total number of photon flux
in the line.)
We note that $F_{\rm K\alpha}$ is consistent with constancy if unity 
abundances are assumed; this is in agreement with simple power law fits.
However, this leaves us with difficult-to-constrain errors, and worse fits
in a $\chi^2$ sense. (With the exception of $F_{\rm K\alpha}$, all other
parameters as e.g. $\Gamma$ shown in Table~\ref{tab2-pexrav} follow similar trends whether 
unity or non-unity abundances are assumed.)  It is clear that degeneracies 
exist and cannot be resolved with 3$-$20~keV {\it RXTE } data - we suspect that 
the dependence on abundance is largely due to the modelling of the  iron edge in these
complex fits.  
Nevertheless, we test this hypothesis by including an edge feature to the simple power law model
of Table~\ref{tab1-pl}.  These results presented in Table~\ref{tab-powithedge} show
that $F_{\rm K\alpha}$ is indeed consistent with constancy and strengthens the 
argument that the behaviour of the iron line as given by the complex model of 
Table~\ref{tab2-pexrav} is complicated with degeneracies.
This and the flux constancy of the iron line is well illustrated in Fig.~\ref{fig7-ratioplts}b
which shows the ratio of the best-fit data against the model of Table~\ref{tab2-pexrav}.
Certainly, the case is strong for a requirement of
supersolar iron abundances and is reflected in the strength of the
iron line.  We discuss this in depth in Lee et al. (1999).

\begin{figure*}
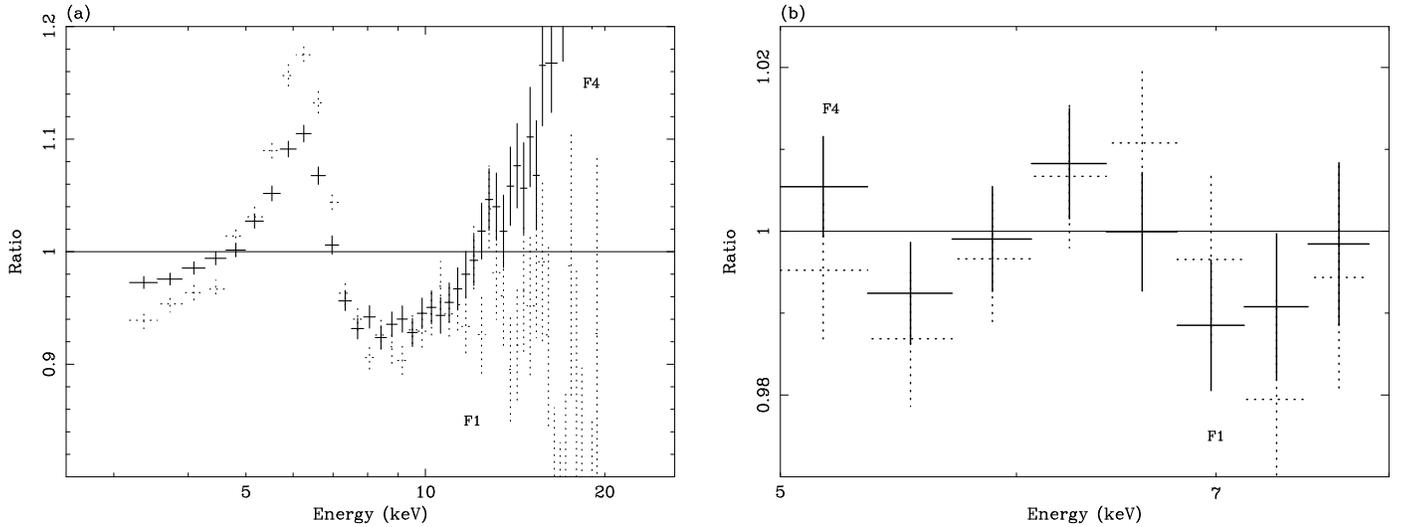

  \hbox{
    \psfig{figure=m762f7a.ps,width=0.5\textwidth,angle=270}
    \hspace{0.5cm}
    \psfig{figure=m762f7b.ps,width=0.5\textwidth,angle=270} }
\caption[h]{(a) Ratio plot of data-to-powerlaw\_model of the iron line and reflection
component for comparison between the extreme flux states {\bf f1} (lowest)
and {\bf f4} (highest). A simple power law is used to fit the data.
(b) Ratio of bestfit data-to-model for the flux states {\bf f1} and {\bf f4} 
illustrates the flux constancy of the iron line.  
The model used is that presented in Table~\ref{tab2-pexrav}. 
Dotted lines represent {\bf f1} and solid lines {\bf f4}.    }
\label{fig7-ratioplts}
\end{figure*}

While $F_{\rm K\alpha}$ apparently decreases with flux from complex fits, we find that the
reflection fraction, and absolute normalization of the reflection component
($Rnorm$) increases with flux (Table~\ref{tab2-pexrav} and Fig.\ref{fig6-refewfe}b).  We define
$Rnorm = A * R$, where $A$ is the power law flux at 1~keV, 
in units of $10^{-3}$ $\ph\cm^{-2}\s^{-1}\keV^{-1}$.  (Fig.~\ref{fig7-ratioplts}a clearly
shows that stronger reflection is present during higher flux states.)
The anticorrelation between $F_{\rm K\alpha}$ and reflection (Fig.~\ref{fig6-refewfe}c)
can be due to an `artificial' effect in which the presence
of a strong reflection spectrum during the high flux states has the
effect of removing part of the flux from the line  in the fits,
thereby resulting in lower observed $F_{\rm K\alpha}$ during the
higher flux states.  This is linked to the strong coupling between
the fit parameters of $\Gamma,\ R, {\rm and}\ abundance$, discussed previously.
There is the possibility that iron becomes
more ionized as the  flux increases which will weaken the observed
line flux. (We discuss ionization scenarios in Section~7.) 
As always we caution the effect of inadequate
spectral resolution  and possibly incomplete models.

Additionally, we find that $W_{\rm K\alpha}$  anticorrelates with 
$R$ (Fig.~\ref{fig6-refewfe}d). This lack of 
proportionality between $R$ and $W_{\rm K\alpha}$ is unexpected in the context of the standard
corona/disk geometry, the implications of which are discussed in
Section~7.  Chiang et al. (1999) find similar results in their
multi-wavelength campaign of NGC5548.
In light of degeneracies associated with complex fits, we primarily present
our results from  simple power law fits.

\subsection{The bright {\it ASCA } and {\it RXTE } flares}
We next investigate in greater detail the time sequences surrounding
the brightest {\it ASCA } and {\it RXTE } flares.  In particular, we are prompted by
the peculiar behaviour of the iron line surrounding the time interval
of  the {\it ASCA } flare as reported by I99 for this time sequence as seen
in the {\it ASCA } data.  
According to I99, there is dramatic change of the line in both profile and intensity 
during this period: interval {\it a} as depicted in Fig.~\ref{fig9-ascaxte-flare}
is marked by an extremely redshifted line profile that is characterized by a
 sharp decline in the line energy at 5.6~keV (far below the 6.4~keV 
rest energy of the line emission) with a red wing that extends to 3.5~keV,
and line intensity that is $\sim$3 times that for the time averaged data.
The transition to time interval {\it b} is marked by an abrupt factor of 2.2 drop in
the averaged 0.6-10~keV count rate, and similar $\sim$2 drop in intensity.  
(We present in this section the behaviour as seen by {\it RXTE }.) 
 Fig.~\ref{fig9-ascaxte-flare}
shows the {\it ASCA } and corresponding {\it RXTE } light curves for this period of
interest.  We note that the two observations are slightly offset from
each other; this will allow for a good assessment of the events
immediately preceding and following this flare event.  Unfortunately
{\it ASCA } data do not exist to coincide with the most prominent {\it RXTE } flare
during the time interval between $2.8 \times 10^5$ and $3.8\times
10^5$ s.

We fit the 3-10~keV data with a number of different models : (1) {\sc power
law + Gaussian}, (2) {\sc (power law + Gaussian)} modified by absorption, (3)
{\sc power law + diskline} and (4) {\sc (power law + diskline)} modified by
absorption.  For the majority of the fits, the {\sc diskline} model by Fabian
et al. (1989) provided the best results; these {\sc power law + diskline}
best fit values are detailed in Table~\ref{tab3-ascaflare} and
Table~\ref{tab4-rxteflare}, respectively for the {\it ASCA } and {\it RXTE } flare events. (We note
however that differences between $\chi^2$ for fits to the different
models are not great : $\Delta \chi^2$ less than 2 for 1 extra
parameter.)  Due to the inadequate resolution of {\it RXTE } in the iron line
band  such that it is insensitive to the details that a diskline model
can provide, we fix {\sc diskline} parameters at best fit {\it ASCA } values
reported by I99 for  this 1997 observation : emissivity $\alpha=-4.1$,
inclination $i=30^\circ$, respectively inner and outer radius
$R_{in}=6.7$, and $R_{out}=24$.   We caution that fixing these values is likely
to be an oversimplification of the true scenario since the line profile
is known to vary on short time scales (e.g. Iwasawa et al. 1996), the impact
of which is not easily assessable given present data quality. Nevertheless, 
we find that it is presently the best option for giving a first order approximation of 
what is likely to be happening. 

\subsubsection{The {\it ASCA } flare}
\begin{figure}
\psfig{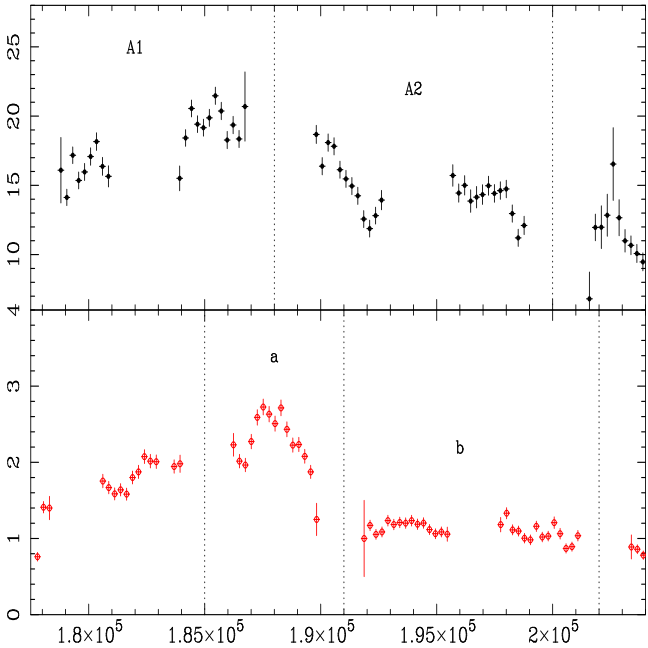}

\caption[h]{{\it RXTE } (top) and {\it ASCA } (bottom) light curves corresponding to the most prominent {\it ASCA } flare. Intervals {\it a} and {\it b} in bottom panel correspond to the flare and post-flare phases discussed in Iwasawa et al. (1999).}
\label{fig9-ascaxte-flare}
\end{figure}

\begin{figure}
\psfig{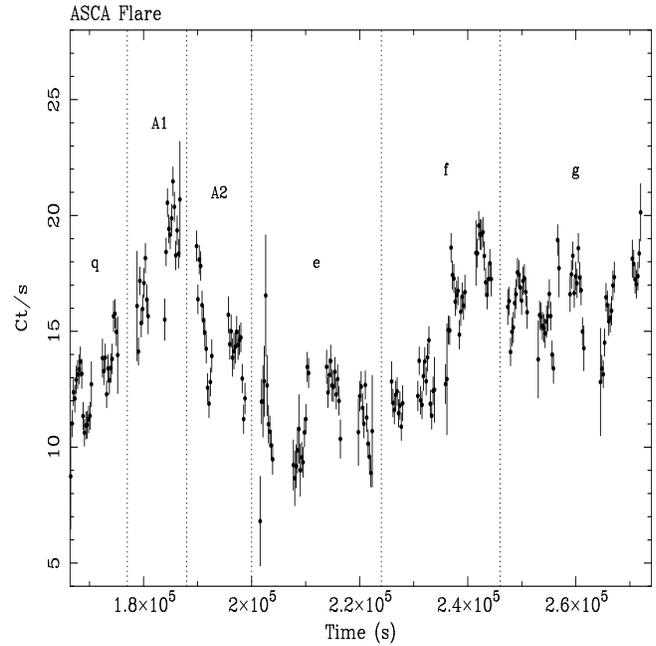}
\caption[h]{{\it RXTE } light curve depicting the intervals of interest preceding and following the most prominent flare {\it a} seen by {\it ASCA }. } 
\label{fig10-ascaflare}
\end{figure}

Figs.~\ref{fig9-ascaxte-flare} \& \ref{fig10-ascaflare} illustrate the {\it RXTE } time intervals corresponding to the
periods surrounding the {\it ASCA } flare, with periods {\bf A1} and {\bf A2}
chosen to respectively correspond to the times immediately preceding
and following the flare `{\it a}' seen in {\it ASCA }.  
Table~\ref{tab3-ascaflare} details fit results for the time intervals depicted in Fig.~\ref{fig10-ascaflare}.

\begin{figure}
\centerline{\psfig{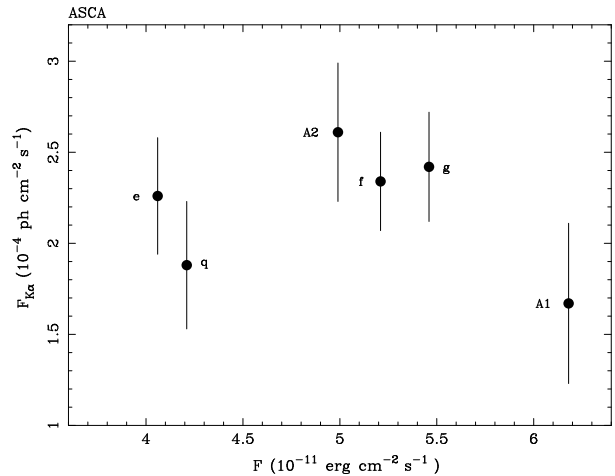}}
\caption[h]{{\it RXTE } light curve depicting the intervals of interest preceding and following the most prominent flare seen by {\it ASCA }.}
\label{fig11-ascaflareplt}
\end{figure}

There are noticeable changes in the intrinsic power law slope throughout
the intervals of interest.  In particular, the most dramatic changes
occur between the intervals {\bf q} and {\bf e}. $\Gamma_{3-10}$
steepens from {\bf q} to {\bf A1} intervals immediately preceding the
flare event seen in {\it ASCA }, which may have the effect of producing the
noticeable decrease in $W_{\rm K\alpha}$.  In the $\sim$ 11~ks
separating the `~pre-~' and `~post-~' flare event (respectively {\bf A1} and
{\bf A2}), $F_{\rm K\alpha}$ increases (from $1.67 \pm 0.44$ to $2.61 \pm 0.38$
\phpcmsqps) and $W_{\rm K\alpha}$ (from $278 \pm 74$ to $543 \pm
79$~eV), nearly double.   This is consistent with I99
findings from {\it ASCA } data for factor of $\sim$ 2 difference
in $F_{\rm K\alpha}$ for flare {\it a}, and post-flare {\it b}
intervals shown in Fig.~\ref{fig9-ascaxte-flare}.  It also appears that $\Gamma_{10-20}$
steepens during this transition, although errors are too large to
definitively make this claim.  Subsequent changes to $\Gamma_{3-10}$
are still noticeable ; $\Gamma_{10-20}$ appears to flatten in {\bf e},
and along with $W_{\rm K\alpha}$ and flux remain fairly constant
through {\bf g}.  Unfortunately, the errors associated with 
$F_{\rm K\alpha}$  are generally too large to make statistically significant
statements about changes in the iron line flux, even though there are 
indications for this (e.g. Fig.~\ref{fig11-ascaflareplt}); notice especially the difference in
$F_{\rm K\alpha}$ between {\bf A1} and {\bf A2}.

\begin{table*}
\begin{center}
\begin{tabular}{|c|c|c|c|c|c|l|}
\multicolumn{7}{c}{\sc Change in $\Gamma$ for time periods previous to and following {\it ASCA } flare } \\

\hline
{\em \rm Interval} & {\em $\rm \Gamma_{3-10}^a$} & {\em $E_{\rm K\alpha}^b$} & {\em $F_{\rm K\alpha}^c$} & {\em $\rm \Gamma_{10-20}^d$} & {\em $W_{\rm K\alpha}^e$ (eV) } & {\em $\rm Flux^f$ }  \\
\hline
\hline
q &  $1.92 \pm 0.04$  & $6.50 \pm 0.15$ & $1.88 \pm 0.35$ & $1.52^{+0.25}_{-0.27}$ & $466 \pm 87$ &  3.25\\
{\bf A1}  & $ 2.05 \pm 0.04$  & $6.37 \pm 0.19$ & $ 1.67 \pm 0.44 $ & $1.53^{+0.24}_{-0.27}$ & $ 278 \pm 74 $ & 4.62  \\
{\bf A2}  & $ 1.99 \pm 0.04$ & $ 6.38 \pm 0.11 $  & $ 2.61 \pm 0.38 $ & $1.91^{+0.28}_{-0.27}$ & $ 543 \pm 79 $ & 3.80 \\
e  & $ 1.85 \pm 0.04$ & $ 6.45 \pm 0.11 $   & $ 2.26 \pm 0.32 $ & $1.66^{+0.24}_{-0.27}$ & $ 566 \pm 80 $ &  3.18\\
f  & $ 1.97 \pm 0.03$ & $ 6.47 \pm 0.09 $  & $ 2.34 \pm 0.27 $ & $1.70^{+0.18}_{-0.18}$ & $ 471 \pm 55$ &  3.98\\
g  & $ 1.96 \pm 0.03$ & $ 6.48 \pm 0.10$  & $ 2.42 \pm 0.30 $ & $1.61^{+0.19}_{-0.18}$ & $ 466 \pm 58 $ &  4.17 \\
\hline

\end{tabular}
\caption{Results are quoted from simple power law fits; the diskline model of Fabian et al. (1989) is used 
to model the line emission.
$^a$ Power-law photon index for 3~keV~$<$~E~$<$ 10~keV.
$^b$ Centroid energy of the iron K$\alpha$ line. 
$^c$ Flux of iron emission line in units of $10^{-4}$ \phpcmsqps .
$^d$ Power-law photon index for 10~keV~$<$~E~$<$~20~keV.  
$^d$ Equivalent width of the iron emission line.  
$^e$ 3-10~keV flux in units of $10^{-11}$ \ergpcmsqps.}

\label{tab3-ascaflare}
\end{center}
\end{table*}

\subsubsection{The {\it RXTE } flare}
\begin{figure}
\psfig{file=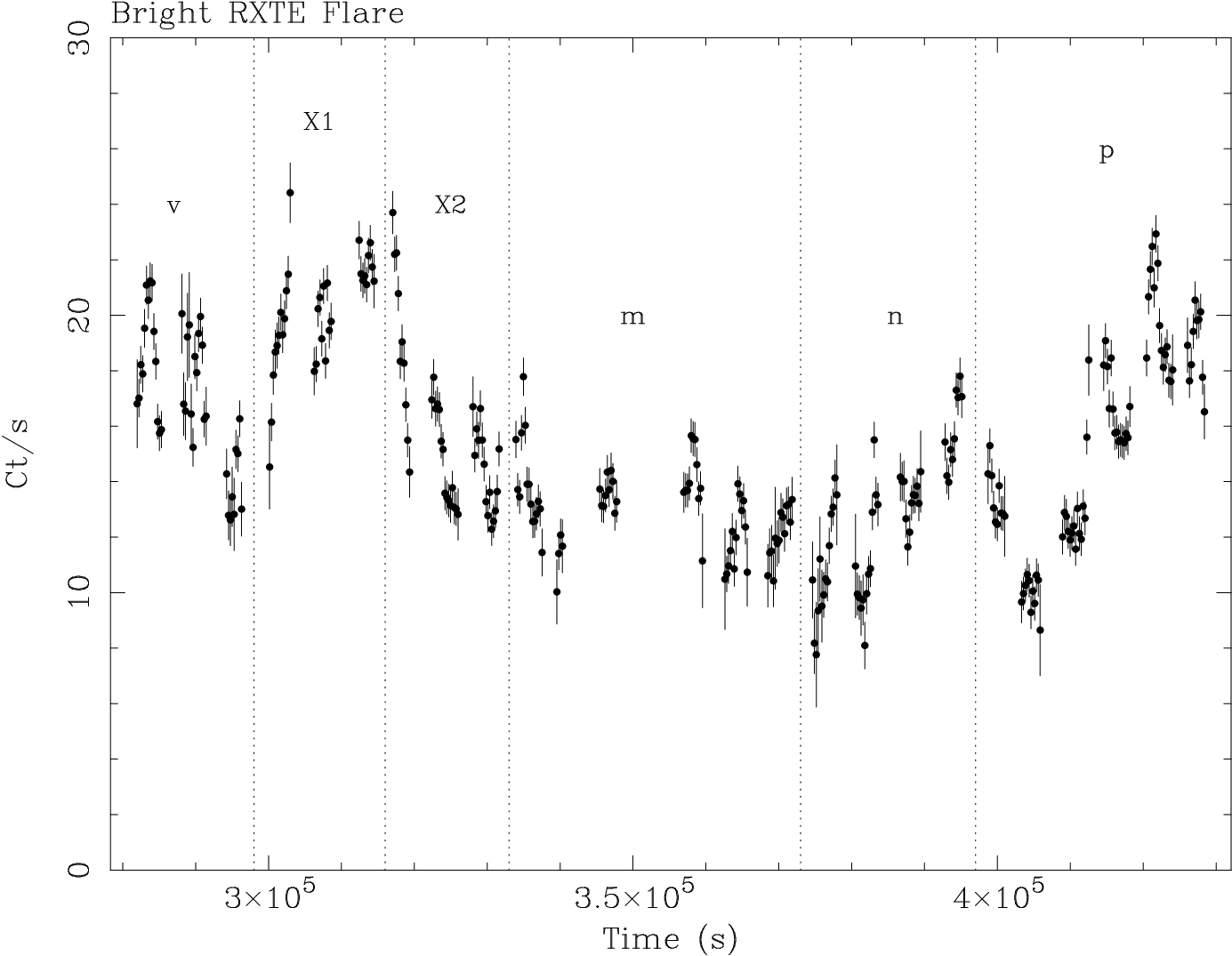,angle=270,width=8.5truecm,height=8.5truecm}

\caption[h]{{\it RXTE } light curve depicting the intervals of interest preceding and following the most prominent flare seen by {\it RXTE }.}
\label{fig12-xteflare}
\end{figure}

\begin{figure}
\centerline{\psfig{figure=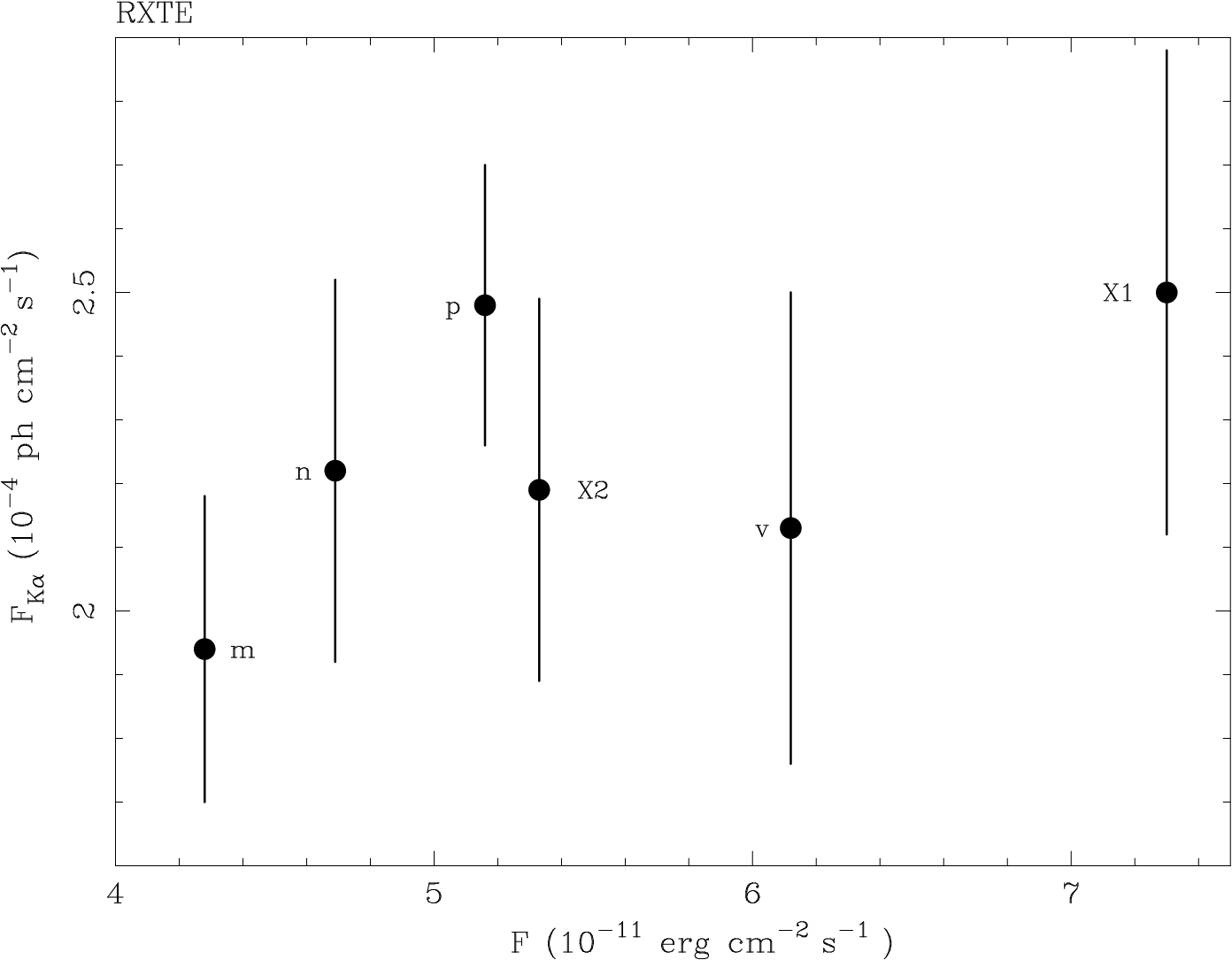,width=0.45\textwidth,angle=270}}
\caption[h]{{\it RXTE } light curve depicting the intervals of interest preceding and following the most prominent flare seen by {\it RXTE} .}
\label{fig13-xteflareplts}
\end{figure}

We next investigate the time intervals surrounding the {\it RXTE } flare (Fig.~\ref{fig12-xteflare}); due
unfortunately to an absence of {\it ASCA } data during this time interval, we
cannot make analogous comparisons.  However, we find that changes to
the  intrinsic power law slope during these intervals follow similar
trends presented for the {\it RXTE } counterpart of the {\it ASCA } flare in the
previous section, although events surrounding this bright {\it RXTE } flare
event appear to be much more erratic and complicated. (We remind the
reader that this is also apparent in hardness ratio comparisons during
this  time interval.) The similarities with the {\it ASCA } flare lie in
observed trends to changes in $\Gamma_{3-10}$. In particular, there is
a  noticeable flattening of $\Gamma_{3-10}$ in the transition between
`~pre-~' and `~post-~' flare states {\bf X1} to {\bf X2}, which continues
through interval {\bf m}, before suddenly steepening in the following
interval {\bf n}.  Unfortunately, errors are such that again we are
unable to make any statements regarding changes to $F_{\rm K\alpha}$
or $\Gamma_{10-20}$.  However, we note that $F_{\rm K\alpha}$ is
generally high for these time intervals associated with this hard {\it RXTE }
flare event.  Table~\ref{tab4-rxteflare} details these results.  As with the {\it ASCA } flare,
Fig.~\ref{fig13-xteflareplts} for the {\it RXTE } flare hints at changes in $F_{\rm K\alpha}$ on 
short time intervals (e.g. {\bf X1} and {\bf m}) 

\begin{table*}
\begin{center}
\begin{tabular}{|c|c|c|c|c|c|l|}
\multicolumn{7}{c}{\sc Change in $\Gamma$ for time periods previous to and following {\it RXTE } flare } \\

\hline
{\em \rm Interval} & {\em $\rm \Gamma_{3-10}^a$} & {\em $E_{\rm K\alpha}^b$} &  {\em $F_{\rm K\alpha}^c$} & {\em $\rm \Gamma_{10-20}^d$} & {\em $W_{\rm K\alpha}^e$ (eV) } & {\em $\rm Flux^f$ }  \\

\hline
\hline
v  & $ 1.97 \pm 0.03$  & $ 6.47 \pm 0.14 $ & $ 2.13 \pm 0.37 $ & $1.74^{+0.22}_{-0.21}$ & $  361 \pm 63  $ &  4.65 \\
{\bf X1}  & $ 2.05 \pm 0.03$  & $ 6.40 \pm 0.12 $ & $ 2.50 \pm 0.38 $ & $1.73^{+0.19}_{-0.18}$ & $  357 \pm 54 $ &  5.46 \\
{\bf X2}  & $ 1.93 \pm 0.03$  & $ 6.35 \pm 0.10 $ & $ 2.19 \pm 0.30 $ & $1.69^{+0.19}_{-0.18}$ & $ 409 \pm 57 $ &  4.09 \\
m  & $ 1.83 \pm 0.03$  & $ 6.49 \pm 0.10 $ & $  1.94 \pm 0.24 $ & $1.57^{+0.16}_{-0.17}$ & $  457 \pm 56 $ &  3.35 \\
n  & $ 1.98 \pm 0.03$  & $ 6.46 \pm 0.11 $ & $  2.22 \pm 0.30 $ & $1.83^{+0.24}_{-0.23}$ & $  497 \pm 68 $ &  3.58 \\
p  & $ 1.94 \pm 0.02$  & $ 6.45 \pm 0.07 $ & $  2.48 \pm 0.22 $ & $1.61^{+0.14}_{-0.14}$ & $  501 \pm 45 $ &  3.96  \\
\hline
\end{tabular}
\caption{Results are quoted from simple power law fits; the diskline model of Fabian et al. (1989) is used 
to model the line emission.
$^a$ Power-law photon index for 3~keV~$<$~E~$<$~10~keV.
$^b$ Centroid energy of iron K$\alpha$ line.
$^c$ Flux of iron emission line in units of $10^{-4}$ \phpcmsqps .
$^d$ Power-law photon index for 10~keV~$<$~E~$<$~20~keV.  
$^e$ Equivalent width of the iron emission line.  
$^f$ 3-10~keV flux in units of $10^{-11}$ \ergpcmsqps.}

\label{tab4-rxteflare}
\end{center}
\end{table*}

\subsection{The intervals {\bf i1} to {\bf i7} }
It is clear from the spectral analysis thus far that 
conditions can alter suddenly and erratically.  In
order to assess whether a more simplified picture exists, we
investigate spectral features of the deep minima in contrast with
flare type events,  using a model that consist of simple power law and
redshifted Gaussian component.  Table~\ref{tab6-intgammachange} confirms the findings of
Section~4.2.1 such that in general, $\Gamma_{3-10}$  tends to be
flatter during the minima in contrast to the flare states.  A close
comparison of $\Gamma_{3-10}$ versus $\Gamma_{10-20}$ for the
differing states suggests that we are largely seeing intrinsic changes
in the power law slope rather than  reflection, although it is likely
that we are seeing contributions from both effects.  Additionally,
ratio plots of data against model using a power law fit show that
there is  a noticeable change in the line flux, profile, as well as
the reflection component, similar to that seen in Fig.~\ref{fig7-ratioplts}a.

\begin{table}
\begin{center}
\begin{tabular}{|c|c|c|c|}
\multicolumn{4}{c}{} \\
{\em \rm Interval} & {\em \rm Exposure} & {\em \rm 2-60~keV} & {\em
 \rm Time range}  \\  & {\em \rm $\rm 10^3$ s} & {\em $\rm ct$ $\rm
 s^{-1}$} & {\em $\rm 10^4 s$ }  \\ \hline
\hline
i1 & 6.74 & 9.59 $\pm$ 0.12  & 3.0-4.8   \\ 
i2 & 9.52 & 15.86 $\pm$ 0.10 & 6.0-8.0   \\ 
i5 & 5.50 & 15.78 $\pm$ 0.14 & 51.0-53.5  \\ 
i6 & 17.65 & 10.50 $\pm$ 0.07 & 54.0-58.0 \\ 
i7 & 19.95  & 16.47 $\pm$ 0.07 & 62.0-66.0  \\
\hline

\end{tabular}
\caption{i1-i7 correspond to the nine intervals of interest for the light curve given in Fig.~\ref{fig1-ascaxteltc}.  The intervals {\bf i1} and {\bf i6} correspond to the deep
minima, whereas  {\bf i2}, {\bf i5}, and {\bf i7} have comparable
fluxes.  Count rates for the intervals are given for the 2-60~keV
energy band of the {\it RXTE } PCA.  }

\label{tab5-intproperties}
\end{center}
\end{table}

\begin{table*}
\begin{center}
\begin{tabular}{|c|c|c|c|c|c|}
\multicolumn{6}{c}{\sc Change in $\Gamma$ for different states of MCG$-$6-30-15} \\
\hline
{\em \rm Interval} & {\em $\rm \Gamma_{3-10}^a$} & {\em $F_{\rm K\alpha}^b$} & {\em $\rm \Gamma_{10-20}^c$} & {\em $\rm W^d$ (eV) } & {\em $\rm Flux ^e$ }  \\
\hline
\hline
i1 & 1.80 $\pm$ 0.06 & $1.79^{+0.59}_{-0.45}$ & 1.77 $\pm$ 0.30& $539^{+178}_{-136}$ & 2.47 \\
i2 & 1.97 $\pm$ 0.03 & $1.73^{+0.35}_{-0.29}$ & 1.59 $\pm$ 0.19 & $322^{+65}_{-54}$ &  3.98 \\
i5 & 1.94 $\pm$ 0.03 & $1.91^{+0.62}_{-0.47}$ & 1.58 $\pm$ 0.16 & $318^{+103}_{-78}$ & 4.17\\
i6 & 1.82 $\pm$ 0.03 & $1.85^{+0.32}_{-0.28}$ & 1.56 $\pm$ 0.16 & $494^{+85}_{-75}$ &  2.68\\
i7 & 1.98 $\pm$ 0.02 & $1.98^{+0.36}_{-0.32}$ & 1.83 $\pm$ 0.14 & $305^{+56}_{-59}$ &  4.33\\
\hline

\end{tabular}
\caption{Results are quoted from simple power law fits; a redshifted Gaussian
is used to model the line emission.
$^a$ Power-law photon index for 3~keV~$<$~E~$<$~10~keV.  $^b$ Flux
of iron emission line in units of $10^{-4}$ \phpcmsqps .  $^c$
Power-law photon index for 10~keV~$<$~E~$<$~20~keV.  $^d$ Equivalent
width of the iron emission line.  $^e$ 3-10~keV flux in units of
$10^{-11}$ \ergpcmsqps.}

\label{tab6-intgammachange}
\end{center}
\end{table*}

\subsection{Summary of spectral findings}
\begin{figure*}
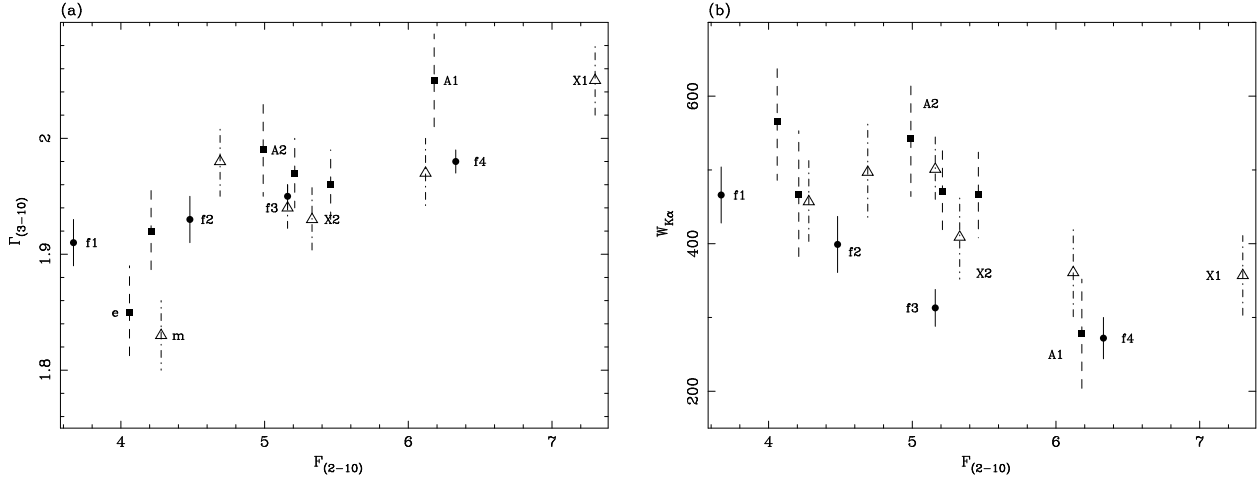

  \hbox{
    \psfig{figure=m762f13a.ps,width=0.45\textwidth,angle=270}
    \hspace{0.5cm}
    \psfig{figure=m762f13b.ps,width=0.45\textwidth,angle=270} }
\caption[h]{2$-$10~keV flux versus (a) $\rm \Gamma_{3-10}$, and (b) equivalent width $W_{\rm K\alpha}$.  Filled circles represent the  different flux states; filled squares and dashed lines represent the {\it ASCA } flare intervals; triangles and dot-dashed lines represent {\it RXTE } flare intervals.  }
\label{fig14-fluxgammaew}
\end{figure*}

We find evidence from flux-correlated studies that $\Gamma_{3-10}$ steepens significantly with flux
($\rm \Delta \Gamma_{3-10} \sim 0.06$ for a doubling of the flux from
{\bf f1} to {\bf f4}; Fig.~\ref{fig14-fluxgammaew}a) while surprisingly, the iron line
strength appears to remain constant (at most differing by $\sim 1.5 \times 10^{-5}$
\phpcmsqps from flux-correlated studies).  Changes to $\Gamma_{10-20}$ ($\rm \Delta \Gamma_{10-20} \sim
0.3$) and $W_{\rm K\alpha}$ ($\Delta W_{\rm K\alpha} \sim 200$~eV) are also
evident and anticorrelate with flux (Fig.~\ref{fig14-fluxgammaew}b for the latter).

A close look at events corresponding to deep minima versus flares reinforces
the finding that changes to the intrinsic power law slope are evident,
with a comparably steeper $\Gamma_{3-10}$ value during the flares.
Figs.~\ref{fig14-fluxgammaew} illustrate that the behaviour of the  intrinsic photon index
and $W_{\rm K\alpha}$ during the flares is consistent  with the
flux-correlated behaviour. 

We find that reflection increases with flux when fitting flux-separated data
with a complex model that includes the reflected spectrum.
Curiously, reflection fraction $R$ anticorrelates with $W_{\rm K\alpha}$; similarly the absolutely 
normalization of the reflection component ($Rnorm$) anticorrelates with $F_{\rm K\alpha}$.
We note that contrary to findings from simple power law fits, $F_{\rm K\alpha}$ is seen to decrease 
with flux, when non-unity abundances are assumed.
We caution however about the large degeneracies associated with complex fits when only the 
3$-$20~keV {\it RXTE } data are considered.  (The case for requiring supersolar abundances
is discussed in Lee et al. 1999). With the exception of the behaviour of $F_{\rm K\alpha}$,
all other parameters as e.g. $\Gamma$ display similar trends using
complex and simple power law fits.   The apparent discrepancy between 
$F_{\rm K\alpha}$ from simple and complex fits in conjunction with 
the large errors associated with $F_{\rm K\alpha}$ would suggest that
an ambiguity exists in measurements of the iron line itself, which {\it RXTE }
is unable to resolve. 

A detailed investigation of the time intervals surrounding the bright
{\it ASCA } and {\it RXTE } flare reveal further complexities.  While changes to
$\Gamma_{3-10}$ are consistent with flux-correlated studies, we find that
there is evidence to suggest that a change in $F_{\rm K\alpha}$ occurs
during the time intervals immediately associated with the flare event;
in other words, tentative evidence for changes to $F_{\rm K\alpha}$ are apparent on short time scales.
In particular,  a large increase in the line flux (e.g {\bf A2}) is
evident in the interval immediately following  the flare;
$W_{\rm K\alpha}$ increases similarly.  It is curious that 
$F_{\rm K\alpha}$ shows a significant increase after the flare rather than
during, and may be an indication that are we witnessing  some type of
response to the flare (e.g. Fig.~\ref{fig11-ascaflareplt}), but also caution that
the evidence is very tentative.  We note also that  $F_{\rm K\alpha}$ is
comparably high ($\sim$ factor of 1.7 increase) during the times
surrounding the {\it RXTE } flare, and times following the  {\it ASCA } flare, in
contrast to values presented in Tables~\ref{tab1-pl}, \ref{tab2-pexrav}, and \ref{tab6-intgammachange}.

\section{Time Lags, Leads, and Reverberation }
Motivated by the temporal findings of e.g. Miyamoto et al. (1988) and Cui
et al. (1997) for Cygnus X-1, we next wish to 
investigate whether the collecting area of {\it RXTE } coupled with this long
observation is sufficient to discern time lags, time leads, and in
particular whether reverberation effects can finally be seen.  We
acknowledge that a good assessment can be hampered by unevenly sampled
data, the nature of  which has been investigated by a number of
workers (e.g. Edelson \& Krolik 1998; Yaqoob et al. 1997) for AGN to
In`t Zand \& Fenimore (1996) for application to gamma-ray bursts.  We
adopt a method similar to that presented by Edelson \& Krolik (1998).
Accordingly, we define the following formalism for our calculations of
the  autocorrelation function (ACF), and cross correlation function
(CCF).

\begin{equation}
CCF(\tau_i) = \frac{1}{M} \sum_{\it i=-n}^{\it n} \frac {y^{\prime}({\it i}) {\it z^{\prime}}({\it i + \tau})}{n_\tau}
\qquad\mbox{for }\qquad \tau \neq 0
\end{equation}
\[
\qquad\mbox{where }\qquad z^{\prime} = z_i - \frac{\sum_{\it i=1}^{\it n}  z_i}{n}
\]
\[
\qquad\mbox{and }\qquad CCF(\tau_0) = \sum_{\it i=1}^{\it n}  (y_i- \bar{y}) (z_i- \bar{z}) = M
\]
The time interval $\tau$ is incremented by n multiples of 64s
($\tau=n\times64$); the subsequent  definition of $n_\tau$ is the
number of bins for which the difference in time between consecutive
time intervals satisfy the present value for $\tau$.  The variables $K$
and $M$ correspond respectively to the ACF and CCF value at $\tau=0$;
these terms are  used in order to normalize the ACF and CCF so that
coherent noise addition at $\tau=0$ is eliminated.

\begin{figure*}
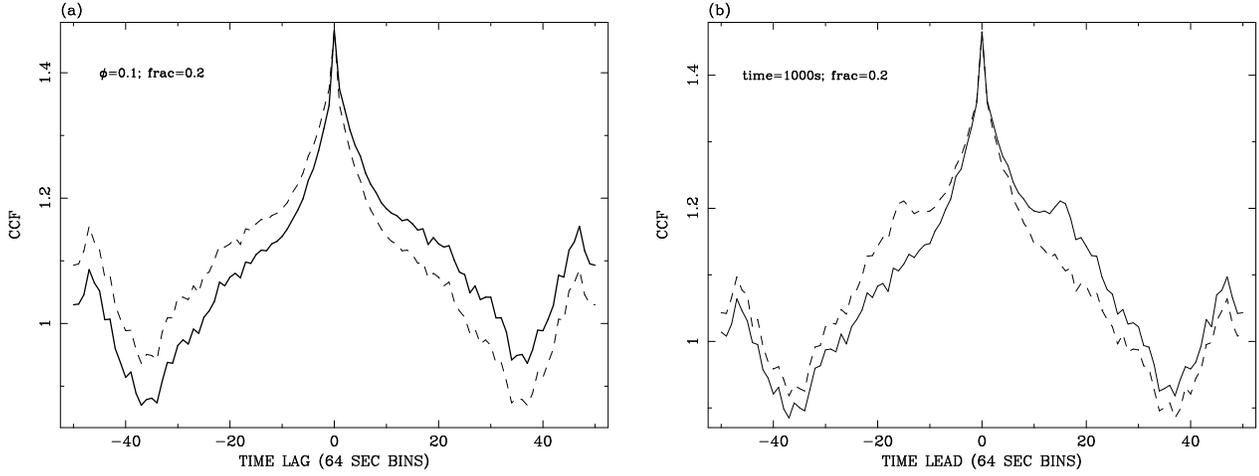

  \hbox{ \psfig{figure=m762f14a.ps,width=0.45\textwidth,angle=270}
    \hspace{0.5cm}
    \psfig{figure=m762f14b.ps,width=0.45\textwidth,angle=270} }

\caption[h]{(a) CCF of simulated light curve with a 20 per cent phase shift of $\sim$ 0.6 radians (solid lines); dashed
lines represent the mirror image to this.
(b) CCF of simulated light curve with a 20 per cent time shift of 1000s (solid lines); dashed lines represent the mirror image to this. }
\label{fig15-simulatedCCF}
\end{figure*}

\begin{figure*}
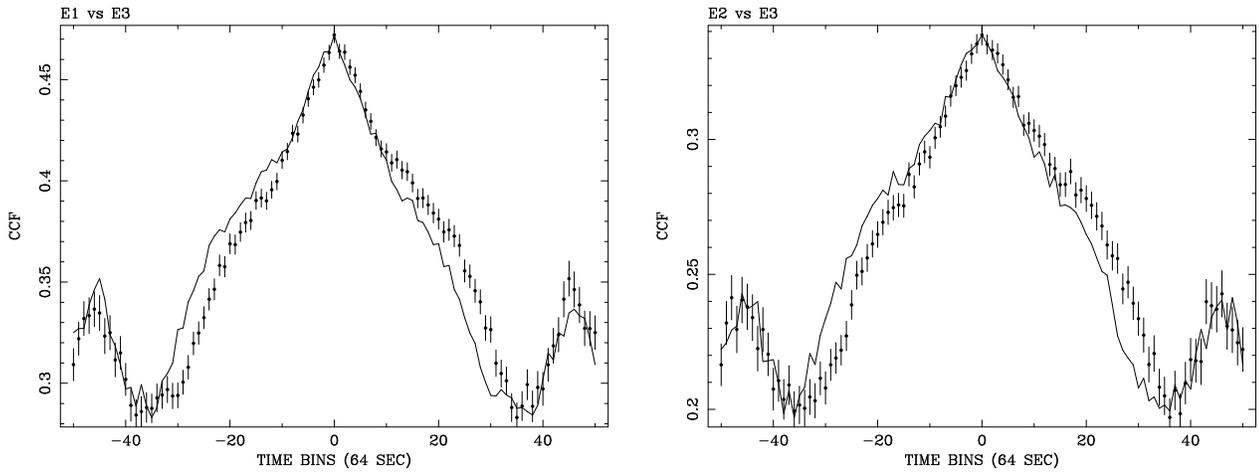

  \hbox{ \psfig{figure=m762f15a.ps,width=0.45\textwidth,angle=270}
    \hspace{0.5cm}
    \psfig{figure=m762f15b.ps,width=0.45\textwidth,angle=270} }

\caption[h]{Data points indicate possible phase lag in (a) CCF between the E1 (2$-$4.5~keV) and E3 (7.5$-$10~keV) light curves and (b) CCF between the E2 (5$-$7~keV) and E3 (7.5$-$10~keV) light curves; superimposed on each CCF is its mirror image as represented by the solid lines. For both cases, the pattern between the CCF and its mirror image is similar to that of Fig.~\ref{fig15-simulatedCCF}a for the simulated data, when a phase shift of $\phi=0.6$~rad is applied.}
\label{fig16-CCFe1e2e3}
\end{figure*}

In order to better understand the nature of our findings shown in
Figs.~\ref{fig16-CCFe1e2e3}-\ref{fig18-e1e2e3e4}, we run our CCF  algorithm on simulated light curves.
The
light curves over which the CCFs are evaluated are  identical except
that one has a specified fraction shifted in phase or time.  In principle,
time- and phase-shifted light curves are identical at certain Fourier frequencies as for example,
for the case in which the power spectrum can be represented by a single sine curve.  However, if the light
curve is the addition of e.g. several sine curves (more representative of
actual physical systems), then a time shift
will have the effect of shifting the entire pattern by the specified time interval,
whereas a phase shift will shift each individual sine curve  along by that phase. The
final outcome of the phase-shifted light curve will consist of the additive components of the
individual sine curves that have been shifted, i.e. the individual frequencies are added
together separately.
The CCF of the simulated phase- and time-shifted light curves are
shown in Figs.~\ref{fig15-simulatedCCF}; solid lines correspond to the CCF, and dashed lines
to the mirror image of this.  We point out that the CCFs show subtle differences.  For the
phase-shifted light curve of Fig.~\ref{fig15-simulatedCCF}a, a constant offset from zero
is seen between the CCF and its mirror image (represented by respectively the
solid and dashed lines).  This is contrasted with the time-shifted light curve in
which the CCF and its mirror image become comparable (i.e. there is no persistent
offset) at $\sim$~30 bins (i.e. $\sim$1900s).
Simulated light curves are generated using
Monte Carlo techniques  for the flux, with power inversely
proportional to frequency, while times are identical to those in the
real data.  We note that the similarity between the phase- and time-shifted data
between 0 and $\sim$10 bins (i.e. $\sim$ 0-640~s) is due to the fact that only 20 per cent of the
light curve has been shifted, with the other 80 per cent of the light curve
that remain identical between the phase-shifted and time-shifted simulations.

Assessment of the CCFs between the upper continuum (E3 : 7.5-10~keV)
and lower continuum (E1 : 3-4.5~keV) and iron line region (E2 :
5-7~keV) shown in Fig.~\ref{fig16-CCFe1e2e3}, reveal evidence for a possible phase
shift.  In a comparison with the CCF of the simulated light curve shown
in Fig.~\ref{fig15-simulatedCCF}a, in which an artificial phase lag of $\phi \sim$ 0.6 rad
is  introduced, we find that a similar trend exists in the actual
data.  This would suggest that $\sim$ 10-20 per cent of the upper
continuum band lags that of the lower energy bands E1 and E2.

We next assess the nature of the (10-20~keV) reflection component with
the other lower energy bands.  In contrast to the previous findings,
the CCFs of the reflection component (E4) with the lower continuum
(E1), and iron line region (E2) points at possible time shifts ($\approxlt$
1000s) as suggested in Fig.~\ref{fig17-e1e2e4}.   Fig.~\ref{fig15-simulatedCCF}b illustrates that the CCFs of
a time shifted light curve is marked by  a non-symmetric bodily shift
of the CCF, whereas this is not the case of the CCF of a light curve
with a phase shift.

Fig.~\ref{fig18-e1e2e3e4} shows that neither a time nor phase shift is seen in a
comparison of the iron line region (E2) with the lower continuum (E1),
and the reflection component (E4) and upper continuum (E3).  CCFs
of the individual energy bands with itself for all energy bands mentioned 
thus far look nearly identical to Figs.~\ref{fig18-e1e2e3e4}.

We note that errors are such that we are unable to make definitive
statements either about a phase or time lag at this time.

\begin{figure*}
  \hbox{ \psfig{figure=m762f16a.ps,width=0.45\textwidth,angle=270}
    \hspace{0.5cm}
    \psfig{figure=m762f16b.ps,width=0.45\textwidth,angle=270} }
\caption[h]{Data points indicate possible time lag in (a) CCF between the E1 (2$-$4.5~keV) and E4 (10$-$20~keV) light curves and (b) CCF between the E2 (5$-$7~keV) and E4 (10$-$20~keV) light curves; superimposed on each CCF is its mirror image as represented by the solid lines. For both cases, the pattern between the CCF and its mirror image is similar to that of Fig.~\ref{fig15-simulatedCCF}b for the simulated data, when a time shift of $\approxlt$ 1000s is applied.}
\label{fig17-e1e2e4}
\end{figure*}

\begin{figure*}
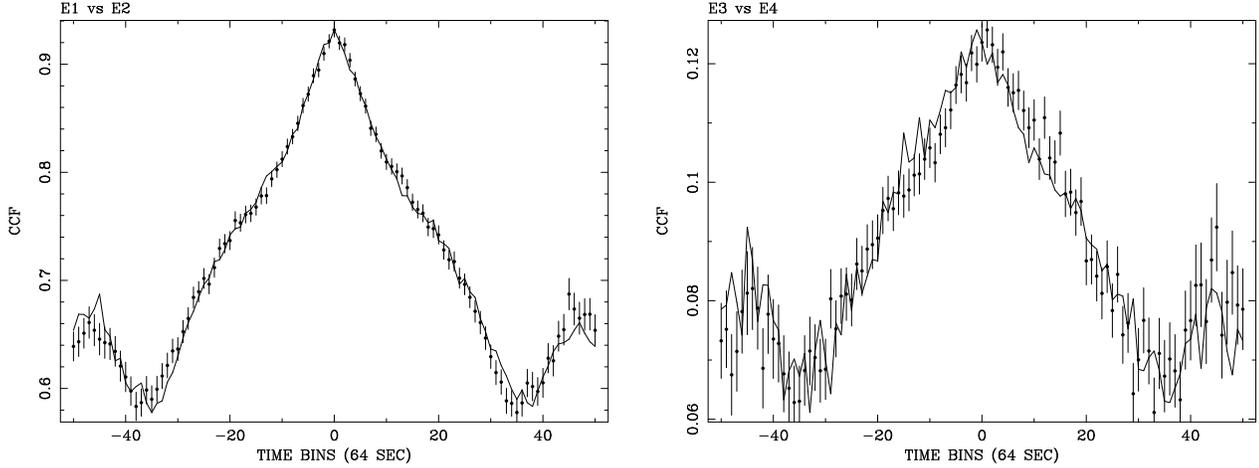

  \hbox{ \psfig{figure=m762f17a.ps,width=0.45\textwidth,angle=270}
    \hspace{0.5cm}
    \psfig{figure=m762f17b.ps,width=0.45\textwidth,angle=270} }
\caption[h]{There are no indications for obvious shifts, either in phase or time for the CCF of the (a) E1 (2$-$4.5~keV) and E2 (5$-$7~keV) light curves, and (b) the  E3 (7.5$-$10~keV) and E4 (10$-$20~keV) light curves.}
\label{fig18-e1e2e3e4}
\end{figure*}

\subsection{Large scale bumps versus small scale flicker}
It is interesting to compare our time-lag results with those of Nowak \& 
Chiang (1999) and Reynolds (1999).   Nowak \& Chiang use similar 
cross-correlation techniques to those employed here to search for time 
lags between the soft band (0.5--1.0\,keV) {\it ASCA } light curve and the 
hard band (8--15\,keV) {\it RXTE } light curve.  They found that the hard 
band lags the soft band by $1.6\pm 0.5$\,ksec.   In this work, we find 
lags $\approxlt 1000$\,s between {\it RXTE } bands E1 (2--4.5\,keV) and E4 
(10--20\,keV).   Noting that thermal Comptonization predicts a time lag 
that varies logarithmically with energy, our results are consistent with 
those of Nowak \& Chiang.

Reynolds (1999) uses an interpolation method to constrain trial transfer
functions linking two given bands.  He found a lag of 50--100\,s between
the 2--4\,keV and 8--15\,keV {\it RXTE } bands, rather smaller than that
found here.  To reconcile these results, one must appreciate that these
methods probe lags at different Fourier frequencies.  The CCF methods
tend to probe lags across a broad spectrum of Fourier frequencies.  Due
to the red nature of the power spectrum, such methods are naturally biased
towards the lower Fourier frequencies.  The method of Reynolds (1999),
instead, probes the higher Fourier frequencies since he uses fairly spiky
trial transfer functions.  Hence, to paraphrase these technical results,
the rapid flickering seems to get transmitted up the observed energy
spectrum with a smaller time lag (by an order of magnitude) than
experienced by the slower variations.

\section{Power spectra and periodicity ? }
While spectral studies of X-ray variability in time sequence may hold
the key to understanding the underlying processes that are responsible
for producing the observed dramatic flux changes,  it is insufficient
for constraining the size of the emitting region in the absence of a
good  understanding of the flare mechanisms.  The line profile
obtained from {\it ASCA } observations suggest that the X-ray emission
originates from $\sim$ 10-20 gravitational radii of the black
hole. This together with a periodic signal can constrain the size of
the emitting region.  Alternatively, we can attempt to estimate the
black hole mass by assessing where the break frequency in the 
PDS occurs (assuming that the break frequency scales with mass).

\begin{figure*}
  \hbox{
    \psfig{figure=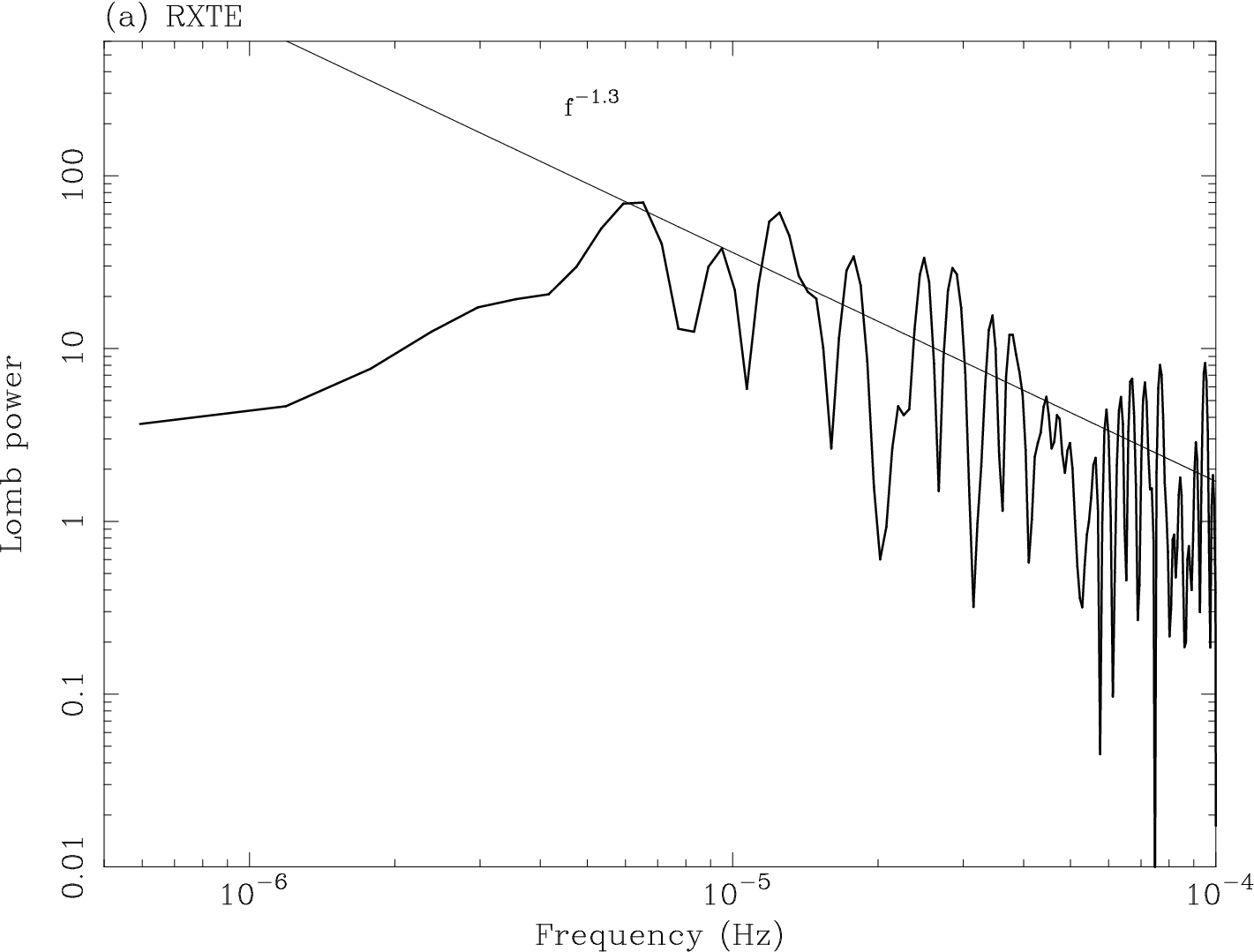,width=0.45\textwidth,angle=270}
    \hspace{0.5cm}
    \psfig{figure=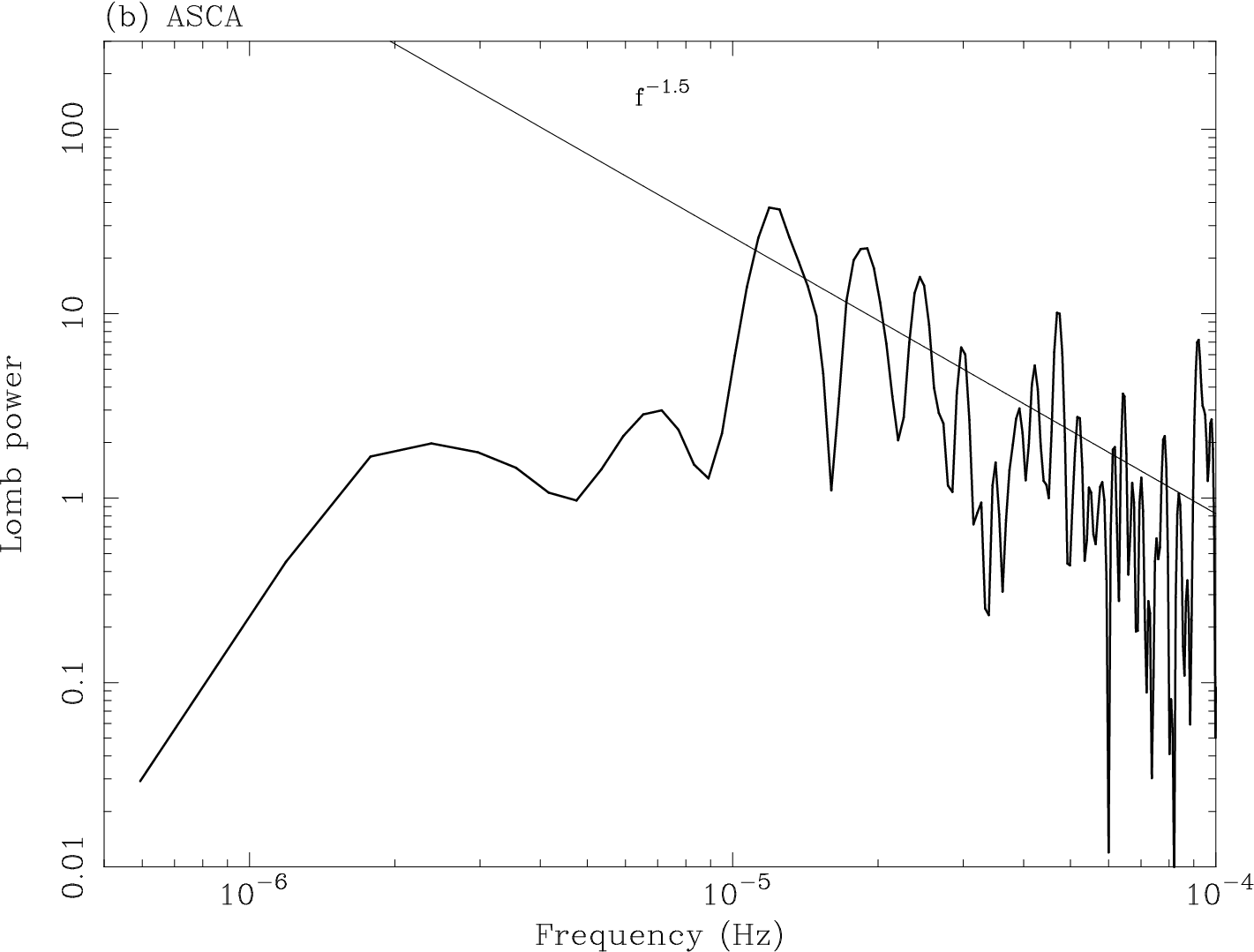,width=0.45\textwidth,angle=270}
    }
\caption[h]{Lomb power as function of frequency for (a) {\it RXTE } 2-60~keV PCA energy band and
(b) {\it ASCA } 0.6-10~keV. A power law model was used to fit the data; best
fit values for the slope are labeled in the figures.}
\label{fig19-lomb}
\end{figure*}

We calculate the power density spectrum using the Lomb-Scargle method
(Lomb 1976; Scargle 1982; Press et al. 1992) appropriate for unevenly
sampled data, and fit a power law slope independently to the {\it RXTE } and
{\it ASCA } data between $10^{-5}$ and $10^{-4}$ Hz.  (We ignore data above
the latter in order  to avoid contamination to the fit from the 96
minute orbital period of {\it RXTE }.)  Figs.~\ref{fig19-lomb} show that respectively,
$f^{-1.3}$ and $f^{-1.5}$ is sufficiently   representative of the {\it RXTE }
and {\it ASCA }  data down to $\sim$ $4-5 \times 10^{-6}$ Hz, where the break
in the power spectrum may occur. 
(We note that this value is only given as a limit
to $f_{br}$ - while there appears to be no additional evidence for a break below $4-5 \times 10^{-6}$~Hz,
the observations are insufficiently long to claim a definitive determination.)
This is consistent with the findings of
Hayashida et al. (1998) for MCG$-$6-30-15.  Since the count rate
throughout these observations remain steadily between 6-24 ct $\rm
s^{-1}$ for {\it RXTE } and 0.5-3 ct $\rm s^{-1}$ for {\it ASCA }, the findings above
would suggest that large scale power does not exist in abundance even
though much shorter time scale variability is highly evident.

We note that Papadakis \& Lawrence (1993) caution against 
standard Fourier analysis techniques for estimating power
spectra, in the form of a bias of the periodogram due to a windowing
effect, in addition to a possible `red noise leak'.  The former is tied
to a concern that the sampling window function can alias power from
frequencies above and below the central frequency $\omega_p$, thereby
distorting the true shape of the power spectrum;  the latter `red
noise leak' effect is concerned with a transfer of power from low to
high frequencies.  (The problem of the  `red noise leak' was first
noted by Deeter \& Boyton (1982) and Deeter (1984).)  However, we
conclude according to subsequent equations (6) and (7) that the
effects of this bias is negligible for the data sets in question.  The
expected value of the periodogram is defined such that :

\begin{equation}
E[I(\omega_p)] = \int^{\frac{\pi}{\Delta T}}_{-\frac{\pi}{\Delta T}}  f(\omega) F_N(\omega - \omega_p) d\omega
\end{equation} 
where $F_N(\omega - \omega_p)$ is the {\it Fejer kernel} (Priestley 1981) which assumes
the shape of the $sine^2$ function for frequencies $\sim \omega_p$.  It follows that the bias is :
\begin{equation}
b(\omega) = E[I(\omega_p)] - f(\omega)
\end{equation}

For lengthy time series (e.g. 400~ks), the mean value of the
periodogram would tend increasingly more towards the true value of the
power spectrum at frequency $\omega_p$, as the  Fejer kernel becomes
increasingly concentrated around this frequency.  In any case,  we only
wish to note that a potential break is seen in the power
spectrum at  $4-5 \times 10^{-6}$ Hz.  (A similar break is noted in this
object from Ginga data by Hayashida et al. 1998.)

\begin{figure}
\centerline{\psfig{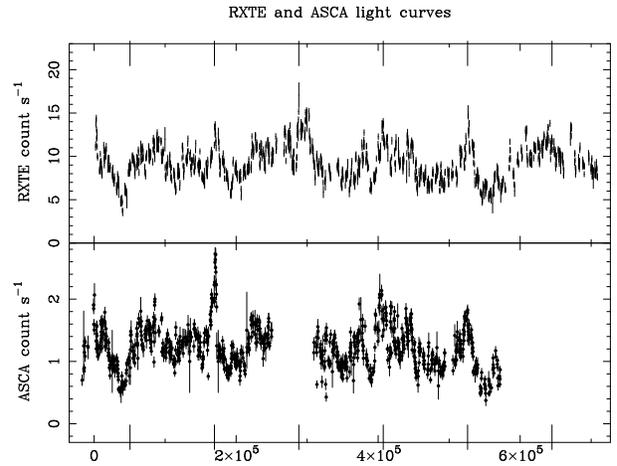}}

\caption[h]{{\it RXTE } (upper) and {\it ASCA } (lower) lightcurves of MCG$-$6-30-15 from
August 1997. Numbering the six tickmarks from left to right, ticks 2 to 5
correspond to small sharp flares seen in the {\it RXTE } data; 2, 4 and 5 are also
seen in the {\it ASCA } data. There is a peak near tick 6 and tick 1 corresponds
to the (jagged) exit from a dip.  Ticks 2 to 5 are regular to within about 1
per cent.}
\label{fig20-33hr}
\end{figure}

We wish additionally to point out an interesting possibility for a 33hr
periodicity.  This is illustrated in Fig.~\ref{fig20-33hr} with 33hr interval
tickmarks superimposed upon the {\it ASCA } and {\it RXTE } light curves.  This has
been determined by taking the mean of the times of the 2 brightest
flares in the {\it RXTE } light curve (the peak of the {\it RXTE } flare {\bf X1}
and {\bf i5}, shown in Fig.~1).    We note that the power of these
peaks is not sufficient to be significant in Lomb-Scargle
power spectra (i.e. the sharp peaks do not carry very much power). 
Accordingly, we have not
attempted to quantify the significance of the peaks, and in part also
because of the red noise nature of the PDS.  We merely point out that
5 out of the  6 tickmarks in the {\it RXTE } light curve with flux $\rm > 14
\thinspace ct \thinspace s^{-1}$ occur in 33~hr intervals, with the
same trend seen in the {\it ASCA } light curve.

\section{Discussion}
There is evidence that the observed spectral variability is complex.
However, all evidence points predominantly to a steepening of the
spectral index with increasing flux. (Other AGNs that have exhibited
spectral variability include NGC7314 Turner 1987;
NGC2992 Turner \& Pounds 1989; NGC4051 Matsuoka et al. 1990; 
NGC3227 Turner \& Pounds 1989, Pounds et al. 1990;  
3C273 Turner et al. 1989, 1990; NGC5548 Nandra et al. 1991, Chiang et al. 1999;
NGC4151 Perola et al. 1986, Yaqoob \& Warwick 1991;)
1H0419-577 Guainazzi et al. 1998.)
 This may be indicative of
changes in the temperature or optical depth of the Comptonizing medium
or of the soft local radiation field.  For instance, we can postulate
that during periods of intense flux, a substantial amount of the hard
X-rays are absorbed by the disk, and thermalized resulting in a source
of soft photons.  These will pass through the corona and Compton cool
it, thereby giving rise to a steeper spectral slope.  However, unless
we fully understand the nature of the competition between coronal
heating and Compton cooling, it is not clear what physics dominates to
give the observed behaviour.

The ejection model of Beloborodov (1998) can explain
the observed relationship between reflection fraction and spectral index 
shown in Table~\ref{tab2-pexrav}.   In this model, observed spectral features are due to 
a non static corona, in which flares are accompanied by plasma ejection
from the active regions.  In other words, the bulk velocity (~$\beta \equiv v/c$~) 
in the flare is $>$ 0.  As the `blobs' move away (are {\it ejected}) from the 
disk, a decrease in reflection is seen as a result of diminished reprocessing 
due to special relativistic beaming of the primary continuum radiation 
away from the disk. The model does not however account for the
constancy of the iron line flux.

We note that the possibility that some of the the iron line and reflection 
may be due to some distant material such as e.g. a torus is ruled out for this 
data based on (1) I96, and I99 findings for a non-constant, and sometimes absent narrow core
seen in long {\it ASCA } observations of MCG$-$6-30-15 in 1994 and 1996, and   
(2) {\it RXTE } findings in this paper that the reflection fraction is seen
to increase with flux.

\subsection{Reflection and the iron line}

A major result from the present observations is the enigmatic
behaviour of the iron line; the inverse proportionality between
$W_{\rm K\alpha}$ and $R$ and of $Rnorm$ with $F_{\rm K\alpha}$. GF91
predict that the bulk of the line arises from fluorescence in
optically thick material.  In this simple reflection picture, we would
expect $R$ and $W_{\rm K\alpha}$ to be proportional to each other provided
(1) the Compton reflection continuum does not dominate the iron line
region, (2) the state of the illuminated region does not change, and
(3) the primary continuum has a fixed spectral shape. We note for the last
point that GF91 point out that differences in the photon index up to
$\Delta \Gamma \sim 0.2$ (as compared to our flux-correlated findings
for $\Delta \Gamma \sim 0.06$) will contribute less than a 10 per cent
effect to $W_{\rm K\alpha}$.

This lack of proportionality between $W_{\rm K\alpha}$ and $R$ (and
$Rnorm$ with $F_{\rm K\alpha}$), in conjunction with an apparently
constant iron line may point at changes to the ionization of the disk
in MCG$-$6-30-15, which is explored further below. We note that
tentative evidence for observed changes in $F_{\rm K\alpha}$ does
exist during time intervals surrounding flare events. Such variability
is clear from the {\it ASCA } analysis (I99). It is possible that 
$F_{\rm K\alpha}$ changes on time scales shorter than is resolvable from
time-averaged spectra; in other words, changes to $F_{\rm K\alpha}$
may only become resolvable during bright flares.

We note that Chiang et al. (1999) find similar results for the
constancy of the iron line and inverse proportionality between
$W_{\rm K\alpha}$ and $R$ in their multi-wavelength campaign of the
Seyfert~1 galaxy NGC5548.

\subsection{A simple model for the observed spectral variations}

We propose the following model in order to explain some of the
enigmatic properties of the observed variability phenomena.  Spectral
variability is no doubt complex, and does not conform to the present
picture of a cold disk geometry for MCG$-$6-30-15. If however the
variable emission from MCG$-$6-30-15 is from a part of the disc which
is more ionized, say with an ionization parameter $\xi\sim 100$ (see
spectra in Ross \& Fabian 1993, Ross et al 1999), then the reflection
continuum will respond to the flux while the iron line does so only
weakly. At that ionization parameter the iron line can be resonantly
scattered by the matter in the surface of the disc and its energy lost
to the Auger process (Ross, Fabian \& Brandt 1996).

The more highly ionized region could either be the innermost regions
of the disc, perhaps within say $6r_{\rm g}$, or the regions directly
beneath the most energetic flares. We note that flux-correlated
changes in the surface density of the disk can lead to changes to
ionization states without incorporating large changes to luminosities
(Young et al. 1999, in preparation).

To illustrate in the context of the {\it RXTE } light curve shown in Fig.~\ref{fig1-ascaxteltc},
assume that the flux below 10 $\rm ct \thinspace s^{-1}$ reflect
physical processes that occur within the radius $6-40 r_g$.  Next,
assume that this is enhanced in the $> 10 \rm \thinspace ct \thinspace s^{-1}$
variability, by flares within $6 r_g$, where the
Auger destruction effect becomes important.  Accordingly, this will lead to
observable changes in reflection (i.e stronger reflection during the
higher flux periods), with minimal changes in $F_{\rm K\alpha}$.

We note that our interpretation that variability largely comes from
within the innermost stable orbit for a Schwarzschild black hole
may be consistent with the scenario for a
very  active corona (and hence strong hard X-ray emission) within $6
r_g$, proposed by Krolik (1999). In this model, magnetic fields within
the radius of marginal stability are strong and amplified
through shearing of their footpoints, which can enhance variability.

\subsection{Implications for Mass Estimates}

\subsubsection{Constraints from spectral studies}
The constancy of the iron line on day-to-day scales suggests that the
timescale for variability (i.e. the observed periods for which 
dramatic flux changes are observed) we are naively probing are much larger (in
the `standard' scenario) than the fluorescing region. In other words,
slow changes would imply on a naive model much larger crossing times
and hence large regions for the crossing times of the continuum. A
light-crossing time of the fluorescing region larger than $\sim$
50~ks (assuming an average radius $\sim 20 r_s$) will lead to an estimate for
the black hole mass $\rm \sim 10^8 M_\odot$. Reynolds (1999) points out
however that the bulge/hole mass relationship of Magorrian et al.
(1998) implies a much lower mass estimate for MCG$-$6-30-15, by an
order of magnitude, of about $\rm \sim 10^7 M_{\odot}$.

In the scenario of the simple model presented above, and given
evidence for short timescale variability of the iron line (here and
I99) as well as the location of the flare line found in I99, the
constancy of the line suggests that the timescale for variability for
which we are probing is much {\it smaller} than the fluorescing region
and reconciles the above mass problem.

\subsubsection{PDS : Analogies with Galactic Black Hole Candidates}
Of further interest is the apparent break in the power spectrum of
MCG$-$6-30-15 seen in both the {\it RXTE } and {\it ASCA } data.  The origin of the
break is not yet known (but see e.g. Edelson \& Nandra 1999; Poutanen \& Fabian 1999;
Kazanas, Hua, \& Titarchuk 1997; and Cui et al. 1997 for possible explanations) , 
but does provide a useful means to determine
the black hole mass, through scaling from similar breaks in the power
spectrum of the famous galactic black hole candidate (GBH) Cygnus X-1,
and other objects like it.

The behaviour of the PDS in MCG$-$6-30-15 is not unlike that of GBHs
in the `low' (hard) state (see e.g. Belloni \& Hasinger 1990; Miyamoto
et al. 1992; van der Klis 1995).  Power law slopes (with form
$f^{\alpha}$) of order $\alpha \sim\ $-1 to -2 are observed  at high
frequencies and flatten to $\sim 0$ at lower frequencies.  If we
bridge the gap between AGNs and GBHs and assume that similar physics
are at play, we can make predictions for the  black hole mass in
MCG$-$6-30-15 (using the values for the cutoff frequencies $f_{br}$)
by a simple scaling relation with Cygnus X-1. Belloni \& Hasinger
(1990) report that $f_{br} \sim$ 0.04-0.4 Hz for Cygnus X-1;  for
MCG$-$6-30-15, we find evidence that $f_{br} \sim 4-5 \times 10^{-6}$
Hz.  The resulting ratio between the 2 cutoff frequencies is $\sim
10^4-10^5$.  Herrero et al. (1995) argue that the black hole mass in
Cygnus X-1 is $\rm \sim 10 M_{\odot}$, which leads us to conclude that
the mass of the black hole in MCG$-$6-30-15 is $\sim 10^5-10^6$,
smaller than anticipated.  

Our mass estimate agrees with that of Hayashida et al. (1998) and 
Nowak \& Chiang (1999) who also used the break frequency and scaling
arguments.  However, such mass estimates should be treated with extreme
caution.  The break frequencies in any one given GBHC can vary by one or
two orders of magnitude depending upon the exact flux/spectral state of
the source.  Given that we do not know how to map AGN spectral states into
analogous GBHC states, the mass estimate derived above (and that of Nowak
\& Chiang 1999) will also be uncertain by up to two orders of magnitude.
Additionally, it is not entirely clear what timescales to identify the 
cutoff frequency with.  A small black hole mass (i.e. $<$ $\rm 2 \times 10^6 M_\odot$)
would also imply the presence of a super-Eddington black hole in MCG$-$6-30-15.

\subsubsection{A possible 33 hour period}
Finally, we address what implications a 33~hr period would have on the
black hole mass in MCG$-$6-30-15. If we make the assumption that this
is the orbital time scale for e.g. a  flare to circumnavigate the
black hole in MCG$-$6-30-15, then we can estimate the mass via  the
relation :

\begin{equation}
M_{BH}(R) = 3.1 \times 10^7 \thinspace t_{orb} \thinspace (\frac{R}{10R_s})^{-1.5} M_{\odot}
\end{equation}
where $R$ is the distance from the center, and $R_s \equiv 2r_g$ is
the Schwarzschild radius (the gravitational radius of the black hole
$r_g \equiv GM/c^2$, and $t_{orb}$ is in days).  

The diskline model constrains $R_{in}$ and $R_{out}$ assuming some
power law emissivity function ($\propto R^{- \alpha}$) that declines
to larger radii. (We note that beyond $r_{out}$ the line emission is
negligible.)   Accordingly, we expect that most of the power  is
concentrated in the inner radii. I96 and I99 constrain using time
averaged {\it ASCA } data  $r_{in} \sim (6.7 \pm 0.8) r_g $ for
MCG$-$6-30-15.  This combined with  $t_{orb} \sim$ 33hr (= 1.375 days)
give a mass for the black hole in  MCG$-$6-30-15 $\rm \sim 2.6 \times 10^7
\thinspace M_{\odot}$.

\section{Conclusion}
We summarize below the spectral and timing results of this paper.  It
is clear  that complicated processes are present, the nature of which
is not obviously apparent, and may prove to be a challenge to present
theoretical models.

\begin{description}
\item \noindent $\bullet$
Hardness ratios reveal that spectral variability may be largely
attributed to changes  in the intrinsic photon index.  In particular,
spectral hardening is observed during  periods of diminished intensity
in comparisons of the (7.5$-$10~keV) upper continuum, (5$-$7~keV) iron line region, and
(10$-$20~keV) reflection hump with the (3$-$4.5~keV) lower continuum. Particularly
hard spectra are noted in a time interval corresponding to $\rm \sim 260~ks$
that begins shortly after the hard {\it RXTE } flare.\\

\item \noindent $\bullet$
We find from flux correlated studies that changes to the photon index
are evident. In particular, $\Gamma_{3-10}$ steepens while
$\Gamma_{10-20}$ flattens with flux;  for a doubling of the  flux,
$\Delta \Gamma_{3-10} \sim 0.06$ and $\Delta \Gamma_{10-20} \sim 0.3$.
This coupled with findings for a constant iron line can contribute to
the reduced fractional variability in the  iron line band noted by
Reynolds (1999).  We note that changes to $\Gamma_{10-20}$ are
significant only with large changes in flux whereas changes in
$\Gamma_{3-10}$ are apparent even with subtle changes in flux. This
point is well illustrated in detailed studies of the time intervals
surrounding the {\it ASCA } and {\it RXTE } flares. Nevertheless, it would appear
that both changes in the intrinsic power law slope (reflected by
changes to $\Gamma_{3-10}$), and reflection (reflected by changes to
$\Gamma_{10-20}$) both contribute in varying degrees to the overall
spectral variability. \\

\item \noindent $\bullet$
We find curiously that the iron line flux is consistent with being
constant over large time intervals on the order of days (but this is 
not the case on much shorter time intervals of order $\approxlt$ 12~ks),
and the equivalent width anitcorrelates with the continuum
flux.  (Observed changes to $F_{\rm K\alpha}$ on short time intervals 
are summarized in the next point.) This may point at evidence 
for a partially ionized disk.\\

\item \noindent $\bullet$
From concentrated studies of the time intervals surrounding the {\it RXTE }
and {\it ASCA } flares, we find tentatively that $F_{\rm K\alpha}$ shows a
noticeable increase after the flare events. (This is less significant
for the periods of the {\it RXTE } flare.) This may be an indication
that we are witnessing some type of response to the flare.  We note
that $F_{\rm K\alpha}$ is comparably high ($\sim$ factor of 1.7
larger) during the times surrounding the {\it RXTE } flare, and times following
the {\it ASCA } flare, in contrast to time averaged analysis of
flux-correlated data. \\

\item \noindent $\bullet$
We find tentative evidence from cross correlation techniques for a
possible phase lag comparable to $\phi \sim 0.6$ between the
(7.5$-$10~keV) upper continuum, and (5$-$7~keV) iron line band and
(3$-$4.5~keV) lower continuum.   \\

\item \noindent $\bullet$
CCFs further reveal possible time lags (time delays $\approxlt$ 1~ks) between
the (10$-$20~keV) reflection hump and iron line band, and reflection
hump with lower continuum.\\

\item \noindent $\bullet$
We report an apparent break (from $f^0$ to $\sim$ $f^{-1.5}$) of
MCG$-$6-30-15 at $\sim 4-5 \times 10^{-6}$~Hz ($\sim 56-70$~hrs) seen
by both {\it ASCA } and {\it RXTE }. Scaling with the mass of the GBH Cygnus X-1 gives a
smaller than expected black hole mass of  $\rm \sim 10^5-10^6
M_{\odot}$ for MCG$-$6-30-15.  However, this is unlikely
to be a proper estimate of the mass for the black hole in MCG$-$6-30-15.
(A black hole mass  $<$ $\rm 2 \times 10^6 M_\odot$ would make the black hole in
MCG$-$6-30-15 super-Eddington.) \\

\item \noindent $\bullet$
We report on the possibility for a 33~hr period seen in both the {\it ASCA }
and {\it RXTE } light curves.  This combined with a value of $\sim 7 r_g$ for
the inner radius, implies black hole mass $\rm \sim 2.6 \times 10^7 \thinspace
M_{\odot}$ for MCG$-$6-30-15. \\

\end{description}

\section*{ACKNOWLEDGEMENTS}
We thank Juri Poutanen for useful discussions about cross correlation
techniques.  We thank all the members of the {\it RXTE } GOF for answering
our inquiries in such a timely manner.  JCL thanks the Isaac Newton
Trust, the Overseas Research Studentship programme (ORS) and the
Cambridge Commonwealth Trust for support.  ACF thanks the Royal
Society for support.   KI and WNB thank PPARC
and NASA {\it RXTE } grant NAG5-6852 for support respectively.
CSR thanks the National  Science Foundation for
support under grant AST9529175, and NASA for support under the Long
Term Space Astrophysics grant NASA-NAG-6337.
CSR also acknowledges support from Hubble Fellowship
grant HF-01113.01-98A awarded by the Space Telescope Institute,
which is operated by the Association of Universities for Research in
Astronomy, Inc., for NASA under contract NAS\,5-26555.

\end{document}